\input harvmac
\let\includefigures=\iftrue
\newfam\black
\includefigures
\input epsf
\def\figin{\epsfcheck\figin}\def\figins{\epsfcheck\figins}
\def\epsfcheck{\ifx\epsfbox\UnDeFiNeD
\message{(NO epsf.tex, FIGURES WILL BE IGNORED)}
\gdef\figin##1{\vskip2in}\gdef\figins##1{\hskip.5in}
\else\message{(FIGURES WILL BE INCLUDED)}%
\gdef\figin##1{##1}\gdef\figins##1{##1}\fi}
\def\DefWarn#1{}
\def\figinsert{\goodbreak\midinsert}
\def\ifig#1#2#3{\DefWarn#1\xdef#1{fig.~\the\figno}
\writedef{#1\leftbracket fig.\noexpand~\the\figno}%
\figinsert\figin{\centerline{#3}}\medskip\centerline{\vbox{\baselineskip12pt
\advance\hsize by -1truein\noindent\footnotefont{\bf Fig.~\the\figno:}
#2}}
\bigskip\endinsert\global\advance\figno by1}
\else
\def\ifig#1#2#3{\xdef#1{fig.~\the\figno}
\writedef{#1\leftbracket fig.\noexpand~\the\figno}%
#2}}
\global\advance\figno by1}
\fi

\def\Tr{{\rm Tr}}

\def\da{{\dot a}}

\def\la{\langle}
\def\ra{\rangle}
\def\al{\alpha}

\def\vac{{\rm vac}}
\def\tN{{\tilde N}}
\def\rmon{{\rm on}}
\def\rmoff{{\rm off}}
\def\rma{{\rm a}}
\def\rmc{{\rm c}}
\def\BMN{{\rm BMN}}
\lref\ZhouMI{
J.~G.~Zhou,
``pp-wave string interactions from string bit model,''
arXiv:hep-th/0208232.
}

\lref\MetsaevBJ{
R.~R.~Metsaev,
``Type IIB Green-Schwarz superstring in plane wave Ramond-Ramond
background,''
Nucl.\ Phys.\ B {\bf 625}, 70 (2002)
[arXiv:hep-th/0112044].
}

\lref\ConstableVQ{
N.~R.~Constable, D.~Z.~Freedman, M.~Headrick and S.~Minwalla,
``Operator mixing and the BMN correspondence,''
JHEP {\bf 0210}, 068 (2002)
[arXiv:hep-th/0209002].
}

\lref\KristjansenBB{
C.~Kristjansen, J.~Plefka, G.~W.~Semenoff and M.~Staudacher,
``A new double-scaling limit of N = 4 super Yang-Mills theory and
PP-wave  strings,''
Nucl.\ Phys.\ B {\bf 643}, 3 (2002)
[arXiv:hep-th/0205033].
}

\lref\LeeVZ{
P.~Lee, S.~Moriyama and J.~w.~Park,
``A note on cubic interactions in pp-wave light cone string field
theory,''
arXiv:hep-th/0209011.
}

\lref\PankiewiczTG{
A.~Pankiewicz and B.~.~Stefanski,
``pp-wave light-cone superstring field theory,''
arXiv:hep-th/0210246.
}

\lref\PankiewiczGS{
A.~Pankiewicz,
``More comments on superstring interactions in the pp-wave background,''
JHEP {\bf 0209}, 056 (2002)
[arXiv:hep-th/0208209].
}

\lref\HeZU{
Y.~H.~He, J.~H.~Schwarz, M.~Spradlin and A.~Volovich,
``Explicit formulas for Neumann coefficients in the plane-wave
geometry,''
arXiv:hep-th/0211198.
}

\lref\MetsaevRE{
R.~R.~Metsaev and A.~A.~Tseytlin,
``Exactly solvable model of superstring in plane wave Ramond-Ramond
        background,''
Phys.\ Rev.\ D {\bf 65}, 126004 (2002)
[arXiv:hep-th/0202109].
}

\lref\BlauNE{
M.~Blau, J.~Figueroa-O'Farrill, C.~Hull and G.~Papadopoulos,
``A new maximally supersymmetric background of IIB superstring
theory,''
JHEP {\bf 0201}, 047 (2002)
[arXiv:hep-th/0110242].
}

\lref\BerensteinJQ{
D.~Berenstein, J.~M.~Maldacena and H.~Nastase,
``Strings in flat space and pp waves from N = 4 super Yang Mills,''
JHEP {\bf 0204}, 013 (2002)
[arXiv:hep-th/0202021].
}

\lref\BeisertBB{
N.~Beisert, C.~Kristjansen, J.~Plefka, G.~W.~Semenoff and M.~Staudacher,
``BMN correlators and operator mixing in N = 4 super Yang-Mills theory,''
arXiv:hep-th/0208178.
}

\lref\SpradlinAR{
M.~Spradlin and A.~Volovich,
``Superstring interactions in a pp-wave background,''
Phys.\ Rev.\ D {\bf 66}, 086004 (2002)
[arXiv:hep-th/0204146].
}

\lref\SpradlinRV{
M.~Spradlin and A.~Volovich,
``Superstring interactions in a pp-wave background. II,''
arXiv:hep-th/0206073.
}

\lref\RoibanXR{
R.~Roiban, M.~Spradlin and A.~Volovich,
``On light-cone SFT contact terms in a plane wave,''
arXiv:hep-th/0211220.
}

\lref\GrossSU{
D.~J.~Gross, A.~Mikhailov and R.~Roiban,
``Operators with large R charge in N = 4 Yang-Mills theory,''
Annals Phys.\  {\bf 301}, 31 (2002)
[arXiv:hep-th/0205066].
}

\lref\SantambrogioSB{
A.~Santambrogio and D.~Zanon,
``Exact anomalous dimensions of N = 4 Yang-Mills operators with large
R  charge,''
Phys.\ Lett.\ B {\bf 545}, 425 (2002)
[arXiv:hep-th/0206079].
}

\lref\GrossMH{
D.~J.~Gross, A.~Mikhailov and R.~Roiban,
``A calculation of the plane wave string Hamiltonian from N = 4
super-Yang-Mills theory,''
arXiv:hep-th/0208231.
}

\lref\GomisWI{
J.~Gomis, S.~Moriyama and J.~w.~Park,
``SYM description of SFT Hamiltonian in a pp-wave background,''
arXiv:hep-th/0210153.
}

\lref\BerensteinSA{
D.~Berenstein and H.~Nastase,
``On lightcone string field theory from super Yang-Mills and
holography,''
arXiv:hep-th/0205048.
}

\lref\PearsonZS{
J.~Pearson, M.~Spradlin, D.~Vaman, H.~Verlinde and A.~Volovich,
``Tracing the string: BMN correspondence at finite J**2/N,''
arXiv:hep-th/0210102.
}

\lref\VamanKA{
D.~Vaman and H.~Verlinde,
``Bit strings from N = 4 gauge theory,''
arXiv:hep-th/0209215.
}

\lref\VerlindeIG{
H.~Verlinde,
``Bits, matrices and 1/N,''
arXiv:hep-th/0206059.
}

\lref\ParnachevKK{
A.~Parnachev and A.~V.~Ryzhov,
``Strings in the near plane wave background and AdS/CFT,''
JHEP {\bf 0210}, 066 (2002)
[arXiv:hep-th/0208010].
}

\lref\ConstableHW{
N.~R.~Constable, D.~Z.~Freedman, M.~Headrick, S.~Minwalla, L.~Motl,
A.~Postnikov and W.~Skiba,
``PP-wave string interactions from perturbative Yang-Mills theory,''
JHEP {\bf 0207}, 017 (2002)
[arXiv:hep-th/0205089].
}

\lref\HokerTZ{
E.~D'Hoker, D.~Z.~Freedman and W.~Skiba,
``Field theory tests for correlators in the AdS/CFT correspondence,''
Phys.\ Rev.\ D {\bf 59}, 045008 (1999)
[arXiv:hep-th/9807098].
}

\lref\BianchiRW{
M.~Bianchi, B.~Eden, G.~Rossi and Y.~S.~Stanev,
``On operator mixing in N = 4 SYM,''
Nucl.\ Phys.\ B {\bf 646}, 69 (2002)
[arXiv:hep-th/0205321].
}

\lref\GreenFU{
M.~B.~Green and J.~H.~Schwarz,
``Superstring Field Theory,''
Nucl.\ Phys.\ B {\bf 243}, 475 (1984).}

\lref\GreenHW{
M.~B.~Green, J.~H.~Schwarz and L.~Brink,
``Superfield Theory Of Type Ii Superstrings,''
Nucl.\ Phys.\ B {\bf 219}, 437 (1983).}

\lref\GursoyFJ{
U.~Gursoy,
``Predictions for pp-wave string amplitudes from perturbative SYM,''
arXiv:hep-th/0212118.
}

\Title{\vbox{\baselineskip12pt\hbox{CALT-68-2424}\hbox{hep-th/0301250}}}
{\vbox{\centerline{SYM
Description of PP-wave String Interactions:} \vskip2pt\centerline
{ Singlet Sector and Arbitrary Impurities}}}

\centerline{Jaume Gomis, Sanefumi Moriyama and Jongwon
Park\footnote{$^\dagger$}
{gomis, moriyama, jongwon@theory.caltech.edu}}
\bigskip\centerline{\it California Institute of Technology 452-48,
Pasadena, CA 91125}
\vskip .3in

\centerline{Abstract}

We study string interactions among string states with arbitrary
impurities 
in the Type IIB plane wave background using string field theory. 
We reproduce all string
amplitudes from gauge theory 
by computing matrix elements of the dilatation
operator in a previously proposed basis of states. 
 A direct correspondence is
found between the string field theory and gauge theory Feynman diagrams.

\Date{1/2003}
\newsec{Introduction}

The solvability \MetsaevBJ\MetsaevRE\ of Type IIB string theory on its
maximally supersymmetric plane wave geometry \BlauNE\ has allowed BMN
\BerensteinJQ\ to represent the {\it free} string spectrum of
excitations around this background in terms of operators of
${\cal N}=4$ SYM. Further checks of this identification were performed
in \GrossSU\ and culminated in \SantambrogioSB , where the exact free
spectrum of the string Hamiltonian was derived from gauge theory
considerations for a class of string states.

The formulation of the duality between {\it interacting} string theory
in the plane wave background and gauge theory has recently been
formulated and tested in \GrossMH\GomisWI (see also \PearsonZS\ for a
complementary description using the string bit model
\VerlindeIG\ZhouMI\VamanKA ). The basic idea is to extend the
classical identification derived from the Penrose limit between the
free string theory Hamiltonian $H_2$ and the gauge theory dilatation
operator\foot{$J$ is the generator of a $U(1)\in SU(4)_R$ subgroup of
the R-symmetry group of ${\cal N}=4 $ SYM.} $\Delta$
\eqn\identifi{{1\over\mu}H_2=\Delta-J,}
to the full interacting theory. In the interacting theory the free
Hamiltonian gets replaced by the interacting Hamiltonian
$H=H_2+g_2H_3+\ldots$, where $g_2$ is the string coupling constant,
and the holographic map proposed in \GrossMH\GomisWI\ reads
\eqn\identifiint{{1\over\mu}H=\Delta-J.}

In \PearsonZS\GomisWI\GrossMH\ a basis\foot{In the next section we
will briefly review how to find the correct basis of gauge theory
states.} of operators in ${\cal N}=4$ SYM was found such that the
${\cal O}(g_2)$ matrix elements of the string Hamiltonian were
reproduced using \identifiint\ from gauge theory computations, which
were initiated in \KristjansenBB\ConstableHW\BeisertBB\ConstableVQ .
The analysis in \PearsonZS\GomisWI\GrossMH\ was restricted to string
states with two different scalar impurities along an ${\bf R}^4$ plane
in the transverse ${\bf R}^8$ directions of the plane wave.
For previous work on string interactions in the  plane wave background,
see
\nref\KiemXN{
Y.~j.~Kiem, Y.~b.~Kim, S.~m.~Lee and J.~m.~Park,
``pp-wave / Yang-Mills correspondence: An explicit check,''
Nucl.\ Phys.\ B {\bf 642}, 389 (2002)
[arXiv:hep-th/0205279].
}
\nref\HuangWF{
M.~x.~Huang,
``Three point functions of N = 4 super Yang Mills from light cone
string  field theory in pp-wave,''
Phys.\ Lett.\ B {\bf 542}, 255 (2002)
[arXiv:hep-th/0205311].
}
\nref\ChuPD{
C.~S.~Chu, V.~V.~Khoze and G.~Travaglini,
``Three-point functions in N = 4 Yang-Mills theory and pp-waves,''
JHEP {\bf 0206}, 011 (2002)
[arXiv:hep-th/0206005].
}
\nref\LeeRM{
P.~Lee, S.~Moriyama and J.~w.~Park,
``Cubic interactions in pp-wave light cone string field theory,''
Phys.\ Rev.\ D {\bf 66}, 085021 (2002)
[arXiv:hep-th/0206065].
}
\nref\ChuQJ{
C.~S.~Chu, V.~V.~Khoze and G.~Travaglini,
``pp-wave string interactions from n-point correlators of BMN
operators,''
JHEP {\bf 0209}, 054 (2002)
[arXiv:hep-th/0206167].
}
\nref\KlebanovMP{
I.~R.~Klebanov, M.~Spradlin and A.~Volovich,
``New effects in gauge theory from pp-wave superstrings,''
Phys.\ Lett.\ B {\bf 548}, 111 (2002)
[arXiv:hep-th/0206221].
}
\nref\HuangYT{
M.~x.~Huang,
``String interactions in pp-wave from N = 4 super Yang Mills,''
Phys.\ Rev.\ D {\bf 66}, 105002 (2002)
[arXiv:hep-th/0206248].
}
\nref\GursoyYY{
U.~Gursoy,
``Vector operators in the BMN correspondence,''
arXiv:hep-th/0208041.
}
\nref\ChuEU{
C.~S.~Chu, V.~V.~Khoze, M.~Petrini, R.~Russo and A.~Tanzini,
``A note on string interaction on the pp-wave background,''
arXiv:hep-th/0208148.
}
\nref\JanikBD{
R.~A.~Janik,
``BMN operators and string field theory,''
Phys.\ Lett.\ B {\bf 549}, 237 (2002)
[arXiv:hep-th/0209263].
}
\nref\BeisertTN{
N.~Beisert,
``BMN operators and superconformal symmetry,''
arXiv:hep-th/0211032.
}
\nref\KiemPB{
Y.~j.~Kiem, Y.~b.~Kim, J.~Park and C.~Ryou,
``Chiral primary cubic interactions from pp-wave supergravity,''
arXiv:hep-th/0211217.
}
\nref\MinahanVE{
J.~A.~Minahan and K.~Zarembo,
``The Bethe-ansatz for N = 4 super Yang-Mills,''
arXiv:hep-th/0212208.
}
\nref\BeisertFF{
N.~Beisert, C.~Kristjansen, J.~Plefka and M.~Staudacher,
``BMN gauge theory as a quantum mechanical system,''
arXiv:hep-th/0212269.
}
\nref\ChuQD{
C.~S.~Chu and V.~V.~Khoze,
``Correspondence between the 3-point BMN correlators and the 3-string
vertex on the pp-wave,''
arXiv:hep-th/0301036.
}
\nref\KloseTW{
T.~Klose,
``Conformal dimensions of two-derivative BMN operators,''
arXiv:hep-th/0301150.
}
$\!\!\!\!\!\!\!\!\!\!\!\!\!\!\!\!\!\!\!\!\!\!\!\!
\!\!\!\!\!\!$\ConstableHW\KiemXN -\KloseTW .

In this paper we compute the ${\cal O}(g_2)$ and ${\cal O}(g_2^2)$
Hamiltonian matrix elements for string states with two {\it identical}
scalar impurities along ${\bf R}^4$ and reproduce them from gauge
theory computations. We find that the matrix elements of the
dilatation operator in the basis described in \GomisWI\ exactly
reproduces the string theory answer. When considering string states
with identical impurities we find that there are new classes of
Feynman diagrams that contribute to the string theory and gauge theory
computations. In this work we find a direct connection between the
Feynman diagrams that appear in the string calculation and the Feynman
diagrams that contribute to the gauge theory matrix elements. 
Roughly, the action of the prefactor in string field theory is
captured by the interaction vertex in gauge theory while the Neumann
matrices are captured by the  sum over all free contractions in
gauge theory. This
correspondence could be an important step in deriving the duality. 
We
then compute the ${\cal O}(g_2)$ Hamiltonian matrix elements for
string states with an {\it arbitrary} number of impurities along
${\bf R}^4$ and exactly reproduce them using gauge theory using
\identifiint\ and the basis of states in \GomisWI,
after identifying gauge theory Feynman diagrams with
corresponding diagrams in string theory.
    These results give strong supporting evidence of the
holographic map \identifiint\ and of  the basis of gauge theory states
proposed in \GomisWI\ as a dual description of string states.

The plan of the rest of the paper is as follows. In section $2$ we
review the holographic map proposed  in \GrossMH\GomisWI\ and review
the basis of states introduced in \GomisWI\ in which to compute gauge
theory quantities.  In section $3$ we consider
the string states and gauge theory operators with two identical scalar
impurities. We perform computations up to ${\cal O}(g_2^2)$ of the string
Hamiltonian matrix elements, emphasizing the extra diagrams that
contribute beyond those that appear when considering string states
with two different scalar impurities. Using the basis change proposed
in \GomisWI\ we exactly reproduce the string theory results from a
gauge theory analysis. In section $4$ we show equivalence between
string theory and gauge theory computations for arbitrary string
states
by identifying string
theory Feynman diagrams with gauge theory Feynman diagrams. We
conclude in section $5$. Appendix $I$, which is outside the main focus
of the paper, contains the ${\cal O}(g_2)$ calculation of a two-impurity
$p$-string state transition into a $p+1$-string state. We find
precise agreement with the gauge theory 
calculation in \GursoyFJ\ once we change to
the basis in \GomisWI. The rest of the appendices summarize the
calculations performed throughout the paper.

\newsec{Review}

As mentioned in the introduction, the proposal for the holographic map
between string theory in the plane wave and ${\cal N}=4$ SYM is
\identifiint . This means, as anticipated by Verlinde \VerlindeIG ,
that all the information is encoded in the matrix of two point
functions of BMN operators
\eqn\teopoint{|x|^{2\Delta_0}\la O_A\bar{O}_B\ra
=G_{AB}+\Gamma_{AB}\ln(x^2\Lambda^2)^{-1},}
where $G_{AB}$ is the inner product metric and $\Gamma_{AB}$ is the
matrix of anomalous dimensions. The proposal \identifiint\  requires
the eigenvalues of $H$ and $\mu(\Delta -J)$ to be the same. However,
as emphasized in \GrossMH\PearsonZS\GomisWI, comparison of matrix
elements of these operators can be achieved in a suitable basis. The
basic principle is to orthonormalize the gauge theory Hilbert space
inner product $G_{AB}$ order by order in $g_2$, which captures
operator mixing\foot{The relevance of mixing in the duality was first
pointed out in \BianchiRW .}
 between BMN operators with different number of
traces. By orthonormalizing, the gauge theory inner product coincides
with the string theory Fock space inner product. The precise mapping
between string theory Fock space states $|s_A\ra$ and gauge theory
orthonormal states $|\tilde{O}_A\ra$ is given by
\eqn\map{
|s_A\ra\rightarrow |\tilde{O}_A\ra=U_{AB}|O_B\ra,\ \hbox{with}\
UGU^\dagger={\bf 1},\ \hbox{and}\
\la s_A|s_B\ra=\la \tilde{O}_A|\tilde{O}_B\ra=\delta_{AB},}
where $|O_A\ra$ are states created by BMN operators. Then, one can
compute the matrix elements of the dilatation operator in the
orthonormal basis and compare with string theory Hamiltonian matrix
elements\foot{As in our previous paper, we will omit the trivial factor on
the right hand side of $(2.3)$  proportional to the classical
dimension $\Delta^0$ from now on.}
\eqn\interact{
{1\over \mu}
\la s_A|H|s_B\ra =\la\tilde{O}_A|(\Delta-J)\tilde{O}_B\ra=
(U\left(\left[\Delta^0-J\right]G+        \Gamma
\right)U^\dagger)_{AB}=n\delta_{AB}+\tilde{\Gamma}_{AB},} 
where $\Gamma$ is the matrix of anomalous dimensions of BMN
operators and $n$ is the number of impurities. 

The change of basis $U$ is, however, not unique. In
\PearsonZS\GomisWI\GrossMH\  a basis was found for which the string
theory matrix elements of string states with two different impurities
was reproduced from gauge theory using \interact. As emphasized in
\GomisWI\ the change of basis is the unique one which orthonormalizes
the gauge theory inner product and leads to the matrix $U$ being
{\it real and symmetric}. We will show that this change of basis is
universal by reproducing the string theory Hamiltonian matrix elements
for arbitrary string states via matrix elements of the dilatation
operator in the universal basis\foot{In our previous work \GomisWI ,
and in the rest of the paper we will refer to this basis as the string
field theory basis.}. In this basis the matrix of anomalous
dimensions was evaluated in \GomisWI\ and are given in terms of the
BMN inner product metric $G$ and matrix of anomalous dimensions
$\Gamma$ as
\eqn\anombasis{\eqalign{
\tilde\Gamma^{(0)}&=\Gamma^{(0)},\cr
\tilde\Gamma^{(1)}&=\Gamma^{(1)}
-{1\over 2}\{G^{(1)},\Gamma^{(0)}\},\cr
\tilde\Gamma^{(2)}&=\Gamma^{(2)}-{1\over2}\{G^{(2)},\Gamma^{(0)}\}
-{1\over2}\{G^{(1)},\Gamma^{(1)}\}
+{3\over8}\bigl\{\bigl(G^{(1)}\bigr)^2,\Gamma^{(0)}\bigr\}
+{1\over4} G^{(1)}\Gamma^{(0)}G^{(1)},}}
where $M^{(s)}$, with $M=\Gamma , G$ or $\tilde\Gamma$, is the $g_2^s$
term in the expansion of $M=M^{(0)}+g_2M^{(1)}+g_2^2M^{(2)}+\ldots$.

\newsec{Correspondence in two impurity singlet sector}

In this section, we study string states and BMN operators with two
real scalar impurities along the same direction in ${\bf R}^4$. Since
$SO(4)$ is a symmetry, we can
decompose two scalar impurity states into ${\bf 4}\otimes{\bf
4}={\bf 1}\oplus
{\bf 6}\oplus {\bf 9}$ irreducible
representations of $SO(4)$, with two repeated
impurities belonging to the singlet. We will consider
states with two impurities in one direction $i\in\{1,2,3,4\}$ instead
of looking at the singlet state and later on extend the analysis to
arbitrary number of impurities.

The single string states we will consider are given by (no sum over $i$):
\eqn\sftstate{\eqalign{ |ii,n\ra &=
\alpha^{i\dagger}_n \alpha^{i\dagger}_{-n} |\hbox{vac}\ra, \cr
|ii,0\ra &= {1\over \sqrt{2}} \alpha^{i\dagger}_0 \alpha^{i\dagger}_0
|\hbox{vac}\ra.}}
As shown by \ParnachevKK\BeisertBB , the corresponding gauge theory operators
when $g_2=0$ are given respectively by
\eqn\singlet{ \eqalign{ {\cal O}^J_{ii,n} &= {1\over \sqrt{JN^{J+2}}}
\left(\sum_{l=0}^J e^{2\pi iln/J}
\Tr\left(\phi_i Z^l \phi_i Z^{J-l}\right)
-\Tr\left(\bar{Z}Z^{J+1}\right)\right), \cr
{\cal O}^J_{ii,0} &= {1\over \sqrt{2JN^{J+2}}}
\left( \sum_{l=0}^J \Tr\left(\phi_i Z^l \phi_i Z^{J-l}\right)
-\Tr\left(\bar{Z}Z^{J+1}\right)\right),}}
without summing over $i$. The extra contribution involving
$\bar{Z}$ is crucial \ParnachevKK\BeisertBB\
        for the existence of the BMN limit,
where  $N,J\rightarrow \infty$, with $g,\ g_2=J^2/N\ \hbox{and}\
\lambda^\prime=g^2N/J^2$ fixed and as
we will see leads to interesting new
effects.

The interaction term $H_3$ couples single string states to
two-string states. These are given by
\eqn\twostr{\eqalign{|ii,m,y\ra\ra
&=\al_m^{i\dagger}\al_{-m}^{i\dagger}|\hbox{vac},y\ra\otimes|
\hbox{vac},1-y\ra,\cr
|ii,0,y\ra\ra
&={1\over
\sqrt{2}}\al_0^{i\dagger}\al_{0}^{i\dagger}|\hbox{vac},y\ra\otimes|
\hbox{vac},1-y\ra,\cr
|ii,y\ra\ra&=\al_0^{i\dagger}|\hbox{vac},y\ra
\otimes\al_0^{i\dagger}|\hbox{vac},1-y\ra,}}
where $0<y<1$ is the fraction of the total momentum carried by the
first string in the two-string state.
    These states are  represented when $g_2=0$ by the
following gauge theory
operators
\eqn\opers{\eqalign{
{\cal T}_{ii,m}^{J,y}
&=:{\cal O}_{ii,m}^{y\cdot J}\cdot{\cal O}^{(1-y)\cdot J}:,\cr
{\cal T}_{ii}^{J,y}
&=:{\cal O}_{i}^{y\cdot J}\cdot{\cal O}_i^{(1-y)\cdot J}:,}}
where $y=J_1/J$ and $1-y=J_2/J$ and
\eqn\gaugethe{\eqalign{
{\cal O}^J &= {1\over \sqrt{JN^{J}}}
\Tr\left(Z^J\right),\cr
{\cal O}_i^J &= {1\over \sqrt{N^{J+1}}}
\Tr\left(\phi_iZ^J\right).}}

We now proceed to describe string interactions among these states
using string field theory and reproduce the results from a gauge
theory analysis.

\subsec{SFT computations}

The proper way to describe string interactions in the light-cone gauge
is by using light-cone string field theory. The Hamiltonian is given
by $H=H_2+g_2H_3+g_2^2H_2^\prime+\ldots$, where $g_2$ is the string
coupling
constant. $H_3$ is the leading interaction and couples an $n$-string
state to an $(n\pm 1)$-string state and $H_2^\prime$ is a contact
term. Following the flat space results in
\GreenHW\GreenFU\ the plane wave vertex $H_3$ has been studied in
\SpradlinAR\SpradlinRV\PankiewiczGS\PankiewiczTG\HeZU .
\bigskip

\noindent
$\bullet${\it The ${\cal O}(g_2)$ Computation}

The properly normalized cubic interaction term in the case of purely
bosonic excitations along ${\bf R}^4$ in the exponential (BMN) basis
of oscillators is given by\foot{We take without loss of generality
$\alpha^\prime p^+_{(3)}=-1, \alpha^\prime p^+_{(1)}=y$ and
$\alpha^\prime p^+_{(2)}=1-y$, where
$0<y<1$. Therefore, $\lambda^\prime=1/\mu^2$. The large
$\mu$ normalization was fixed in \PearsonZS\GomisWI\ by comparison
with a field theory amplitude.}
\eqn\Hone{{1\over\mu}|H_3\ra=-{y(1-y)\over 2}{P|V\ra,}}
where $p^+_{(r)}$ is the length of string $r$ and $P$ is the prefactor
\eqn\Prefa{
P=\sum_{r=1}^3\sum_{n=-\infty}^\infty
{\omega_{n(r)}\over\mu
p^+_{(r)}\alpha^\prime}\al^{i\dagger}_{n(r)}\al^i_{-n(r)},}
with $\omega_{n(r)}=\sqrt{(\mu p^+_{(r)}\al')^2+n^2}$ and\foot{Here we
omit the overall $p^+$ conservation factor,
$|p^+_{(3)}|\delta(p^+_{(1)}+p^+_{(2)}+p^+_{(3)})$.}
\eqn\V{|V\ra=\exp\biggl({1\over2}\sum_{r,s=1}^3
\sum_{m,n=-\infty}^\infty
\al_{m(r)}^{i\dagger}{\tilde
N}^{(rs)}_{mn}\al_{n(s)}^{i\dagger}\biggr)|\hbox{vac}\ra.}

We now compute the matrix elements between single string and
two-string states. It is convenient to introduce Feynman rules to
evaluate these amplitudes, specially in later sections when we
consider arbitrary impurities. They are given by:

\bigskip

\noindent
\eqn\Feynmm{\eqalign{
(r,m)\;\hbox{---------------}\;(s,n)\qquad
&\Longleftrightarrow\qquad\tN^{(rs)}_{m,n},\cr
(r,m)\;\hbox{---------------}
\!\!\!\!\!\!\!\!\!\!\!\!\!\!\!\!\!\!\times
\;\;\;\;\;\;\;\,(s,n)\qquad
&\Longleftrightarrow\qquad\left[{\omega_{m(r)}\over
\mu p^+_{(r)}\alpha^\prime}+{\omega_{n(s)}\over
\mu p^+_{(s)}\alpha^\prime}\right]\tN^{(rs)}_{m,-n},}}

\bigskip

\noindent
where $r,s\in\{1,2,3\}$ label the string and $m,n$ label the
worldsheet momentum of the oscillator.
Then,  the Neumann matrix $\tN^{(rs)}_{m,n}$ is the propagator between
oscillators $\alpha_{m(r)}$ and  $\alpha_{n(s)}$. We can eliminate the
prefactor $P$
in \Hone\ by sequentially commuting it through the external states
oscillators,
which has the effect of reversing the sign of the worldsheet momentum
of the oscillator which $P$ is acting on. After elimination of the
prefactor, we are left
with contractions between external states oscillators.
    The $\times$ symbol
in the vertex \Feynmm\ signifies the total effect of commuting the
prefactor $P$
in \Hone\ through both oscillators and their contraction.

Using these Feynman rules and the following symmetry relations
    satisfied by
    the Neumann matrices
\eqn\sym{\tN^{(rs)}_{m,n}=\tN^{(sr)}_{n,m},\qquad
\tN^{(rs)}_{m,n}=\tN^{(rs)}_{-m,-n},}
we can now evaluate any Hamiltonian matrix element using combinatorics
of Feynman diagrams. In the case of two identical impurities, the
amplitudes
are given by:
\eqn\honecal{\eqalign{
{1\over \mu}\la ii,n|H_3|jj,m,y\ra\ra
=&-{y(1-y)\over 2}\biggl[\delta_{ij}
4\tN^{(13)}_{m,n}\tN^{(13)}_{m,-n}
\biggl({\omega_{m(1)}\over\mu y}
-{\omega_{n(3)}\over\mu}\biggr)\cr
&\qquad\qquad+2\tN^{(33)}_{n,-n}\tN^{(11)}_{m,m}
{\omega_{m(1)}\over\mu y}
-2\tN^{(33)}_{n,n}\tN^{(11)}_{m,-m}
{\omega_{n(3)}\over\mu}\biggr],\cr
{1\over \mu}\la ii,n|H_3|jj,y\ra\ra
=&-{y(1-y)\over 2}\biggl[
\delta_{ij}4\tN^{(13)}_{0,n}\tN^{(23)}_{0,n}
\biggl(1-{\omega_{n(3)}\over\mu}\biggr)\cr
&\qquad\qquad+2\tN^{(33)}_{n,-n}\tN^{(12)}_{0,0}
-2\tN^{(33)}_{n,n}\tN^{(12)}_{0,0}
{\omega_{n(3)}\over\mu}\biggr].}}

We note that there are Feynman diagrams in which the identical
impurities in a given string are connected via Neumann matrices
involving only that string. Such contributions are absent when
considering strings with different impurities due to the $SO(8)$
invariance of the Neumann matrices. We can evaluate the expression in
the large\foot{We summarize the large $\mu$ expansion of the Neumann
matrices in the Appendix A.} $\mu$ limit, which corresponds to
the perturbative gauge theory regime. Even though $\tN^{(11)}$,
$\tN^{(12)}$ and $\tN^{(33)}$ are suppressed by $1/\mu$ as compared to
$\tN^{(13)}$, the self-contraction contributions are of the same order
as the contractions between different strings due to cancellations in
the contribution of contractions between different strings. The large
$\mu$ expressions are given by\foot{In light-cone string field theory
the canonical
normalization of states is the usual delta function normalization $\la
s_A^\prime|s_B^\prime\ra=|p_A^+| \delta(p_A^++p_B^+)=J_A
\delta_{J_A,J_B}$, so that
$|s_A^\prime\ra=\sqrt{J_A}|s_A\ra$. Therefore, when comparing string
field theory
results with
gauge theory results we  have to take into account this
normalization factor and the overall $\delta$-function
in \V, since gauge theory states have unit  norm
\BerensteinSA\ConstableHW , so we divide the string theory answer
\honecal\ by
$\sqrt{Jy(1-y)}$.}
\eqn\largemuexp{\eqalign{
{1\over\mu}\la ii,n|H_3|jj,m,y\ra\ra&=\delta_{ij}
\left(\tilde\Gamma^{(1)}_{n,my}+\tilde\Gamma^{(1)}_{-n,my}\right)
-{1\over 2}\Gamma^{(1)}_{n,0y},\cr
{1\over\mu}\la ii,n|H_3|jj,y\ra\ra&=\delta_{ij}
\left(\tilde\Gamma^{(1)}_{n,y}+\tilde\Gamma^{(1)}_{-n,y}\right)
-{1\over 2}\Gamma^{(1)}_{n,y},}}
where
\eqn\expressions{\eqalign{
\tilde{\Gamma}^{(1)}_{n,my}
&=\lambda^\prime{\sqrt{1-y}\over\sqrt{Jy}}
{\sin^2(\pi ny)\over 2\pi^2},\cr
\tilde{\Gamma}^{(1)}_{n,y}
&=-\lambda^\prime{1\over\sqrt{J}}
{\sin^2(\pi ny)\over 2\pi^2}.}}
$\Gamma^{(1)}_{n,0y}$ and $\Gamma^{(1)}_{n,y}$ are defined in Appendix
B and as we shall see have a direct gauge theory origin.
The splitting of the first term in \largemuexp\
into two identical contributions is convenient when comparing with the
gauge theory analysis in the next subsection.

The first contribution in \largemuexp\ is twice as large as compared
to the answer one gets when considering string states with two
different impurities \PearsonZS\GomisWI . The reason is that
there are twice as many
ways of contracting impurities among different strings. This is
reproduced in the gauge theory
computation because the scalar impurities have two ways of contracting
when they are both the same. The last term in \largemuexp\ are due to
self-contractions and only appear when two impurities are repeated. In
the  gauge theory computation in next subsection
these extra contractions are due to the extra diagrams
that one gets when considering the operators \singlet\opers. The new
contractions in string field theory correspond to gauge theory diagrams
involving ${\bar Z}$ and diagrams coupling all four scalar impurities. In
section $4$ the connection between gauge theory diagrams and string
field theory diagrams will be made explicit.
\bigskip

\noindent
$\bullet${\it The ${\cal O}(g_2^2)$ Computation}

We now consider the ${\cal O}(g_2^2)$ matrix elements between single
string states, that is, the contact term contribution. We will also
reproduce this result from gauge theory considerations.

The single string contact term in the plane wave geometry
has been recently analyzed in \RoibanXR . It is constructed from the
plane wave dynamical
supersymmetry generators via $H_2'=\{Q_3,{\bar Q}_3\}$, where $Q_3$
is the leading $g_2$ correction to the free supercharge. In \RoibanXR\
it was shown that by considering the contact term contribution for two
different impurity  string states the gauge theory
results in the orthonormal basis of \PearsonZS\GomisWI\GrossMH\ could
be reproduced if one truncated the intermediate states to the two
impurity sector. We will perform a similar calculation for string
states with two identical impurities using the same
truncation and reproduce these results from gauge theory in the next
subsection.  Understanding more precisely why the truncation works is
an important open problem.

The intermediate two impurity states that contribute are given by
\eqn\inter{\eqalign{|j,m,y,1\ra\ra&=\al_{m}^{j\dagger}
{1\over\sqrt{2}}(b_{m}^{d\dagger}-ie(m)b_{-m}^{d\dagger})
|\vac,y\ra\otimes|\vac,1-y\ra,\cr
|j,0,y,1\ra\ra&=\al_{0}^{j\dagger}b_{0}^{d\dagger}
|\vac,y\ra\otimes|\vac,1-y\ra,\cr
|j,0,y,1\ra\ra'&=\al_{0}^{j\dagger}|\vac,y\ra\otimes
b_{0}^{d\dagger}|\vac,1-y\ra,}}
and $|j,m,y,2\ra\ra(')$ defined by changing the string on which the
operators act. The $b$ oscillators are the fermionic oscillators.
Using the expression in \PankiewiczTG\ for the supersymmetry charge
$Q$ we can calculate its matrix elements in the large $\mu$
limit\foot{The zero mode contribution vanishes in the large $\mu$
regime.}(see Appendix C for details)
\eqn\SUSYelement{\eqalign{&Q_{n,m(s)}
=\la ii,n|Q_\da|j,m,y,s\ra\ra\cr
&\quad\simeq\sqrt{1+\mu\al k}\delta_{ij}u^i_{abc\da}
\delta^{abcd}_{1234}{Y_{m(s)}\over\sqrt{2}}
\left[{\tilde F}^-_{(3)-n}{\tilde N}^{(s3)}_{m,n}
+{\tilde F}^-_{(3)n}{\tilde N}^{(s3)}_{m,-n}\right],}}
for $s=1,2$.
Therefore the ${\cal O}(g_2^2)$ Hamiltonian matrix element
in the  case of two impurities in
the same direction is given by
\eqn\contactsame{
\la ii,n|H_2^\prime|jj,m\ra=\delta_{ij}\int_0^1
{dy\over y(1-y)}\sum_{s=1}^2
\sum_{l=-\infty}^\infty Q_{n,l(s)}Q_{m,l(s)}^*.}
Performing the relevant sums and integral one arrives at
the final result(see Appendix D):
\eqn\contact{\la ii,n|H_2^\prime|jj,m\ra=\delta_{ij}
{1\over 16\pi^2}(B_{n,m}+B_{n,-m}).}

The result in \contact\ has an extra term as compared to the
calculation for two different impurities, which is identical to the
first one except for the sign of the worldsheet momentum.

In this subsection we have calculated the Hamiltonian matrix elements
using string field theory  up to ${\cal O}(g_2^2)$. We now turn to the
gauge theory
analysis.

\subsec{Gauge theory computations}

The BMN operators with two identical scalar impurities \singlet\opers\
are
insensitive to the sign of the worldsheet momentum since
${\cal O}^J_{ii,n}={\cal O}^J_{ii,-n}$ and
${\cal T}^{J,y}_{ii,m}={\cal T}^{J,y}_{ii,-m}$, so we will
consider without loss of generality $n,m
\geq 0$. Moreover, the BPS double trace operator ${\cal
T}^{J,y}_{ii}$ is invariant under $y\rightarrow 1-y$, so we can
restrict to $0< y\leq 1/2$.

As explained in section 2, in order to compute string interactions
from gauge theory we must compute the matrix of two point functions of
BMN operators ${\cal
O}^J_{ii,n},\ {\cal T}^{J,y}_{ii,p}$ and ${\cal
T}^{J,y}_{ii}$. The relevant inner product metric and matrix of
anomalous dimensions can be extracted from \BeisertBB. They are given
by\foot{We have summarized in the Appendix B the explicit expressions for
the matrix elements.}:
\eqn\G{\eqalign{ G={\bf 1}+&g_2\,\delta_{ij}
\pmatrix{0&C_{n,qz}+C_{-n,qz}&2C_{n,z}\cr
C_{py,m}+C_{py,-m} & 0 & 0\cr 2C_{y,m} & 0 & 0\cr}\cr
\noalign{\vskip3pt}
+&g_2^2\,\delta_{ij}\pmatrix{M^1_{n,m}+M^1_{n,-m}&0&0\cr
0&\la?\ra&\la?\ra\cr 0&\la?\ra&\la?\ra\cr},}}
and
\eqn\anomalous{\eqalign{
&{\Gamma}=\delta_{ij}
\pmatrix{\lambda'n^2\delta_{nm}&0&0\cr
0&\lambda'{p^2\over y^2}\delta_{p,q}\delta_{y,z}&0\cr 0&0&0\cr}\cr
\noalign{\vskip3pt}
&\!+g_2\pmatrix{0&\delta_{ij}
\bigl(\Gamma^{(1)}_{n,qz}+\Gamma^{(1)}_{-n,qz}\bigr)
-{1\over2}\Gamma^{(1)}_{n,0z}
&2\delta_{ij}\Gamma^{(1)}_{n,z}-{1\over2}\Gamma^{(1)}_{n,z}\cr
\delta_{ij}\bigl(\Gamma^{(1)}_{py,m}+\Gamma^{(1)}_{py,-m}\bigr)
-{1\over2}\Gamma^{(1)}_{0y,m}&0&0\cr
2\delta_{ij}\Gamma^{(1)}_{y,m}-{1\over2}\Gamma^{(1)}_{y,m}&0&0\cr}\cr
\noalign{\vskip3pt}
&\!+g_2^2\pmatrix{\delta_{ij}
\bigl(\Gamma^{(2)}_{n,m}+\Gamma^{(2)}_{n,-m}\bigr)
-{1\over 16\pi^2}{\cal D}^1_{n,m}&0&0\cr
0&\la ?\ra&\la ?\ra\cr
0&\la ?\ra&\la ?\ra\cr},}}
where $\la ?\ra$ denotes matrix elements that have not yet been
computed.
We note that whenever the worldsheet momentum index in \G\anomalous\
vanishes,
that we must divide the matrix element by
$\sqrt{2}$. Likewise when both operators have vanishing momentum we
must divide that matrix element by 2. These extra factors arise from
our normalization of the operators in \singlet\opers\ which differ
from those in \BeisertBB. In this way we
get an orthonormal inner product for $n,m\ge 0$.

The inner product metric can be computed in   the free theory while
the matrix of anomalous dimensions comes with a power of
$\lambda^\prime$ from evaluating one loop graphs.
In the free theory the $\bar{Z}$ portion of
the gauge theory operators \singlet\opers\ does not couple to the
terms in  \singlet\opers\ without the $\bar{Z}$. Moreover, the
diagrams involving only ${\bar Z}$ are suppressed
by a power of $1/J$ with respect to the leading contribution, which
only involves the part of the operator with the two impurities(terms
without
${\bar Z}$). Therefore,  in
the computation of the mixing matrix the extra term in the operators
\singlet\opers\ does not contribute in the BMN limit,
    so that at any order in $g_2$ the inner product metric can be
calculated neglecting the ${\bar Z}$ term. It then follows that  there
are twice as many contributions in the inner
product of \singlet\opers\ as compared to the case of two different
impurities.
     This is easy to understand since there are
now twice as many ways of contracting the impurities and they
    come with the opposite sign of the worldsheet momentum. An analogous
phenomena occurs when extending the analysis to arbitrary number of
impurities.

The matrix of
anomalous dimensions also has twice as many contributions of the type
appearing for different impurities. These gauge theory Feynman
diagrams can be identified  in the string field theory calculation with
contractions involving impurities living in different strings.
However,
there is an extra term arising from vertices involving ${\bar
Z}$ in  \singlet\opers\ and the coupling of all scalar
impurities\foot{The quartic scalar coupling denotes the effective
interaction after taking into account self-energy and  gluon exchange
diagrams\BeisertBB.} (see Fig. 1).
\ifig\newwww{The new diagrams. The thick lines represent the
impurities or $\bar{Z}$ 
while the thin lines denote $Z$. The first line is for diagrams
involving ${\cal T}^{J,y}_{ii,m}$ while the second is for diagrams
with ${\cal T}^{J,y}_{ii}$.}{\epsfxsize5in\epsfbox{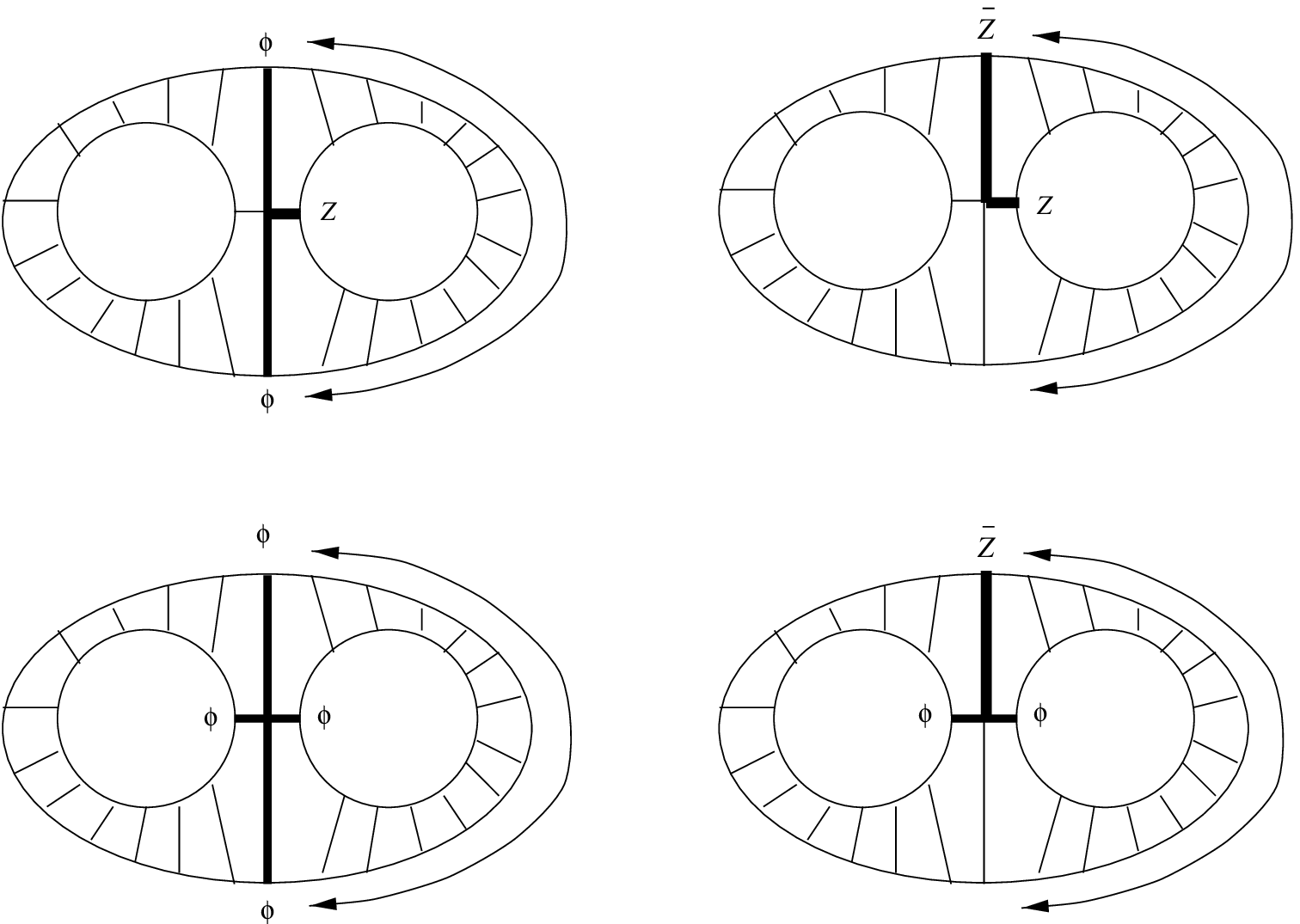}}

 These extra Feynman diagrams can be identified in the
string field theory calculation with contractions of impurities living
on the same string as can be inferred by looking at \largemuexp .

We can now test the holographic correspondence \interact . Using the
formula for the matrix of anomalous dimensions in the orthonormal
basis in terms of $G$ and $\Gamma$ we find:
\eqn\tildeGamma{ \eqalign{ \tilde\Gamma^{(1)}_{ii;n,jj;my}
&=\delta_{ij}\left(\tilde\Gamma^{(1)}_{n,my}+
\tilde\Gamma^{(1)}_{-n,my}\right)
-{1\over2}\Gamma^{(1)}_{n,0y},\cr
\tilde\Gamma^{(1)}_{ii;n,jj;y}
&=\delta_{ij}\left(\tilde\Gamma^{(1)}_{n,y}
+\tilde\Gamma^{(1)}_{-n,y}\right)
-{1\over2}\Gamma^{(1)}_{n,y},}}
where terms with $ \tilde{\Gamma} $ on the right hand side come from
the usual Feynman diagrams present also for two different impurities
and the last term comes from new diagrams only present when two
impurities are the same. By comparing with the string field theory
calculation \largemuexp\ we find precise agreement.

We now proceed to computing the matrix elements of the mostly single
trace operators\foot{In order to compute the matrix elements of the
mostly double trace operators to this order, we would need to know the
expressions for $\la ?\ra$.} to order $g_2^2$. Using \anombasis\ we
find after some computation\foot{For the detailed computation, see
Appendix E.}
\eqn\tildeGammasecond{  \tilde\Gamma^{(2)}_{ii;n,jj;m}  =
\delta_{ij}\left(\tilde\Gamma^{(2)}_{n,m}+\tilde\Gamma^{(2)}_
{n,-m}\right)+
\delta \tilde\Gamma^{(2)}_{ii;n,jj;m},}
where
\eqn\formuladiff{
\tilde\Gamma^{(2)}_{n,m}={1\over 16\pi^2}B_{n,m},}
is the result obtained for different impurity operators \GomisWI\ and
$\delta\tilde\Gamma^{(2)}_{ii;n,jj;m}$ are the new contributions only
arising for identical impurity operators. They are given by\foot{We
use the notation $\delta\hbox{A}$ for the new contributions to $A$ due
to having identical impurities.}
\eqn\DeltaGammasecond{\delta \tilde\Gamma^{(2)}_{ii;n,jj;m}=
\delta\Gamma^{(2)}_{ii;n,jj;m}
-{1\over2}\{G^{(1)},\delta\Gamma^{(1)}\}_{ii;n,jj;m},}
since as explained above only the matrix of anomalous dimensions
receives genuine new contributions while the inner product
contributions have the same form as in the case of different
impurities.
   From \anomalous\ we read
\eqn\DeltaGammasecondfirst{\eqalign{
\delta\Gamma^{(2)}_{ii;n,jj;m}&=-{1\over16\pi^2}{\cal D}^1_{n,m},\cr
\delta\Gamma^{(1)}_{ii;n,jj;my}=\delta\Gamma^{(1)}_{ii;my,jj;n}
&=-{1\over2}\Gamma^{(1)}_{n,0y},\cr
\delta\Gamma^{(1)}_{ii;n,jj;y}=\delta\Gamma^{(1)}_{ii;y,jj;n}
&=-{1\over2}\Gamma^{(1)}_{n,y}.}}
After some computation one finds (see Appendix E for details)
\eqn\GGamma{\{G^{(1)},\delta\Gamma^{(1)}\}_{ii;n,jj;m}
=-{1\over8\pi^2}{\cal D}^1_{n,m},}
giving us the simple result:
\eqn\final{\delta\tilde\Gamma^{(2)}_{ii;n,jj;m}=0.}
Hence, the final expression is
\eqn\tildeGammatwoiijj{{\tilde\Gamma}^{(2)}_{ii;n,jj;m}
=\delta_{ij}{1\over 16\pi^2}(B_{n,m}+B_{n,-m}),}
which exactly matches the ${\cal O}(g_2^2)$ contact term contribution
in the string field theory calculation \contact.

We now turn to the analysis of arbitrary string states.

\newsec{Generalization to arbitrary impurities}

Thus far we have analyzed the correspondence for string states with
two impurities. In this section we construct a proof that shows
the equivalence between the string theory and gauge theory
computations for an arbitrary number of impurities. The idea is
to find a direct link between the Feynman diagrams of string theory
and the Feynman diagrams of gauge theory, so that the equality between
string theory and gauge theory for
arbitrary states follows diagram by diagram. We first outline the
strategy of the proof and then give the explicit details of the string
theory and gauge theory computation.

Let's first consider which diagrams in string theory contribute to
leading order in the $1/\mu$ expansion, which is of  ${\cal
O}(1/\mu^2)$. These diagrams will
have corresponding contributions in the one  loop -- which
is  of ${\cal O}(\lambda^\prime)$ -- gauge theory computation. We
consider matrix elements between single string states and two-string
states with $n$ impurities each, that is impurity
preserving\foot{Impurity non-preserving processes
are inherently non-perturbative \ConstableHW .} processes.
    The impurities can be
distributed at will among the four directions in ${\bf R}^4$.

As explained in section $3$, in order to compute  ${\cal O}(g_2)$
Hamiltonian matrix elements, we must commute the prefactor \Prefa\ of
the cubic
vertex  \Hone\ through all the impurities. These gives us a sum of
$2n$ terms with $2n$ oscillators each
in which the sign of the worldsheet momentum of one of the
oscillators is reserved.  Each term now can be calculated using the
Feynman rules \Feynmm .
Each diagram  is multiplied by the frequency of
the oscillator whose worldsheet momentum is reversed when commuting
through the prefactor.  Now, given the $SO(8)$ invariance of
the string field theory vertex \V, the oscillators in
different directions in ${\bf R}^4$ completely decouple, so we can
concentrate on the case in which all the impurities are in one
direction. The final answer for arbitrary string states is just the
product of the contribution along each of the ${\bf R}^4$ directions.

We can now classify Feynman diagrams in terms of the number
of self-contractions (propagators)
in the single string state, that is the number of
${\tilde
N}^{(33)}$'s.  It is clear that to ${\cal O}(1/\mu^2)$ there can be
{\it at
most} one self-contraction. Since we are looking at impurity
preserving processes, a self-contraction ${\tilde
N}^{(33)}$ always is accompanied by a self-contraction in the
two-string state of the type ${\tilde
N}^{(rs)}$, where $r,s$ is either $1$ or $2$. Since   ${\tilde
N}^{(33)}$ and ${\tilde
N}^{(rs)}$ are of  ${\cal O}(1/\mu)$, we can have at most one
self-contraction to leading order in the $1/\mu$ expansion. This
simple observation greatly diminishes the Feynman diagrams that need
to be considered. We now
study the two possibilities.

Let us consider first the case in which there are no
self-contractions. In this case all impurities in the single string
are contracted with impurities of the two-string state, so the result
is the product of Neumann matrices of the type ${\tilde N}^{(r3)},
(r=1,2)$, where  ${\tilde N}^{(r3)}\simeq {\cal O}(1)$. In any of the
$2n$ terms one gets after commuting the prefactor through the
oscillators there is precisely one oscillator with reversed worldsheet
momentum. This oscillator can now contract with any oscillator in the
single string state or two-string state depending on whether the
reversed oscillator belongs to the two-string or single string state.
For each such contraction there is a corresponding one in which the
sign of the worldsheet momentum of the two oscillators involved in
the contraction is
reversed\foot{This appears from the term one gets after commuting the
prefactor through the other oscillator.}. The combination of these two
contractions we represent by the vertex
$(r,m)\;\hbox{---------------}
\!\!\!\!\!\!\!\!\!\!\!\!\!\!\!\!\!\!\times\;\;\;\;\;\;\;\,(3,l)$
in \Feynmm, where $\times$ signifies the action  of the
prefactor on the oscillators $\alpha_{m(r)}$ and $\alpha_{l(3)}$.
These two terms combine to yield an expression
of ${\cal O}(1/\mu^2)$ due to the leading cancellation of the energy
difference
$\biggl({\displaystyle\omega_{m(r)}\over\displaystyle\mu p^+_{(r)}}
-{\displaystyle\omega_{l(3)}\over\displaystyle\mu}\biggr)$
of these two oscillators in the large $\mu$ limit. Therefore, this
class of diagrams yields an expression given by the  product of $n$
Neumann matrices of the type $\tN^{(r3)}$ for $r=1$ or $2$ times the
energy difference between one oscillator in the single string state and
one oscillator in the two-string state.

We now consider the case with one self-contraction on the single
string state. As mentioned above, and due to the impurity conservation
condition,
this self-contraction is always accompanied by a self-contraction on
the two-string state. Therefore we have a contribution of the form
${\tilde N}^{(33)}\cdot{\tilde N}^{(rs)}$, where $r,s$ is $1$ or $2$,
which is already of order ${\cal O}(1/\mu^2)$. There are now two
possibilities to be considered. Either any of the oscillators involved
in the self-contraction have their worldsheet momentum reversed due to
action of the prefactor or they don't. If they do not, then there is a
contraction connecting the single string state with the two-string
state involving the oscillator with the worldsheet momentum
reversed. Just as in the previous case of no self-contractions, such
diagram always comes accompanied with another one in which the sign of
the worldsheet momentum is reversed on both oscillators involved
in the contraction, yielding the
vertex
$(r,m)\;\hbox{---------------}
\!\!\!\!\!\!\!\!\!\!\!\!\!\!\!\!\!\!\times
\;\;\;\;\;\;\;\,(3,l)$
for $r=1$ or $2$.
Therefore, in this case, the diagram is proportional to
$\biggl({\displaystyle\omega_{m(r)}\over\displaystyle\mu p^+_{(r)}}
-{\displaystyle\omega_{l(3)}\over\displaystyle\mu}\biggr)\cdot
{\tilde N}^{(33)}\cdot{\tilde N}^{(rs)}\simeq{\cal O}(1/\mu^4)$,
so it does not contribute to the leading order result. Therefore, in
the case of one self-contraction the only possibility left is the case
in which the self-contractions involve one oscillator which has the
worldsheet momentum reversed due to the prefactor, so that only
diagrams with the vertex
$(3,m)\;\hbox{---------------}\!\!\!\!\!\!\!\!\!\!\!\!\!\!\!\!\!\!
\times\;\;\;\;\;\;\;\,(3,l)$
or $(r,m)\;\hbox{---------------}\!\!\!\!\!\!\!\!\!\!\!\!\!\!\!\!\!\!
\times\;\;\;\;\;\;\;\,(s,l)$
for $r,s=1$ or $2$ contribute to the leading order result.

{}From now on, let us focus on a particular Feynman diagram and show
agreement between the string field theory and gauge theory
computation.
The string states with $n$ impurities that we need to consider are
given by\foot{The arbitrary phase of the state is determined by
comparison with gauge theory.}
\eqn\nimpuritystates{\eqalign{
|(d_i,n_i)\ra&=i^n\delta_{\sum_i n_i,0}
\prod_i{\alpha_{n_i}^{d_i}}^\dagger
|\hbox{vac}\ra,\cr
|(e_i,p_i);{\cal I}_1,{\cal I}_2;y\ra\ra
&=i^n\delta_{\sum_{j\in {\cal I}_1}p_j,0}
\delta_{\sum_{k\in {\cal I}_2} p_k,0}
\prod_{j\in {\cal I}_1}{\al_{p_j}^{e_j}}^\dagger
|\hbox{vac},y\ra\otimes
\prod_{k\in {\cal I}_2}{\al_{p_k}^{e_k}}^\dagger
|\hbox{vac},1-y\ra,}}
where the $\delta$-functions impose the familiar level matching
condition.
The corresponding level-matched gauge theory operators are given
by\foot{Here are using a simplified notation for the operators, their
precise description is given in Appendix G and H.}:
\eqn\nimpurityBMN{\eqalign{
{\cal O}^J_{(d_i,n_i)}&={1\over \sqrt{JN^{J+n}}}\!
\sum_{0\leq l_1,\cdots,l_n \leq J}\!\!
\Tr\left(Z\ldots Z{\phi_{d_{1}}\over \sqrt{J}}
    Z\ldots Z {\phi_{d_{2}}\over \sqrt{J}}Z \ldots
      Z{\phi_{d_{n}}\over \sqrt{J}}Z\ldots Z\right)\prod_{i=1}^n
t_{i}^{l_i}\cr
&\quad+{\rm terms}\ {\rm involving}\ {\bar Z}
\quad\qquad\qquad{\rm with }\;\;\;\;\sum_{i=1}^n  n_i=0,\cr
{\cal T}^{J,y}_{(e_i,p_i);{\cal I}_1,{\cal I}_2} &=
:{\cal O}^{y\cdot J}_{(e_j,p_j)_{j\in {\cal I}_1}}\cdot
{\cal O}^{(1-y)\cdot J}_{(e_k,p_k)_{k\in {\cal I}_2}}: \qquad {\rm
with}\;\;\;\;
\sum_{j\in {\cal I}_1}p_j=\sum_{k\in {\cal I}_2}  p_k=0,\cr}}
The labels $d_i,e_i \in \{1,2,3,4\}$ denote the direction along ${\bf
R}^4$ of a string oscillator and the corresponding gauge theory
impurity, and  $n_i , p_i \in {\bf Z}$ are their worldsheet momenta
where $t_i=\exp(2\pi in_i/J)$, and
$s_j=\exp(2\pi ip_j/J_1)$ for $j\in{\cal I}_1$ and
$s_k=\exp(2\pi ip_k/J_2)$ for $k\in{\cal I}_2$.
Also we explicitly assign a factor of $1/\sqrt{J}$ to each impurity 
which aids in keeping  track of factors of $J$ during the computation.
${\cal I}_1$ and ${\cal I}_2$ is a
partition of the index set $\{1,\cdots,n\}$, which describes a
particular way of distributing the $n$ impurities among 
string/trace 1 and 2 respectively.

Let us explain the gauge theory
computation of the two-point function of single-trace and double-trace
   BMN operators defined above and exhibit analogies with the string
theory computation.
At one loop order, that is to ${\cal O}(\lambda^\prime)$, we can have
at most a quartic interaction\foot{As shown in
\HokerTZ\ConstableHW\BeisertBB\ the other possible interactions
cancel among themselves due to supersymmetry.}
 vertex, coupling four fields, 
with two of them contracted with the $in$-operator and the other two with
the $out$-operator. There are three kinds of interaction vertices
depending on how far the two fields in the same operator are
separated:
\ifig\Seminearest{The three classes of interaction
vertices.}{\epsfxsize5in\epsfbox{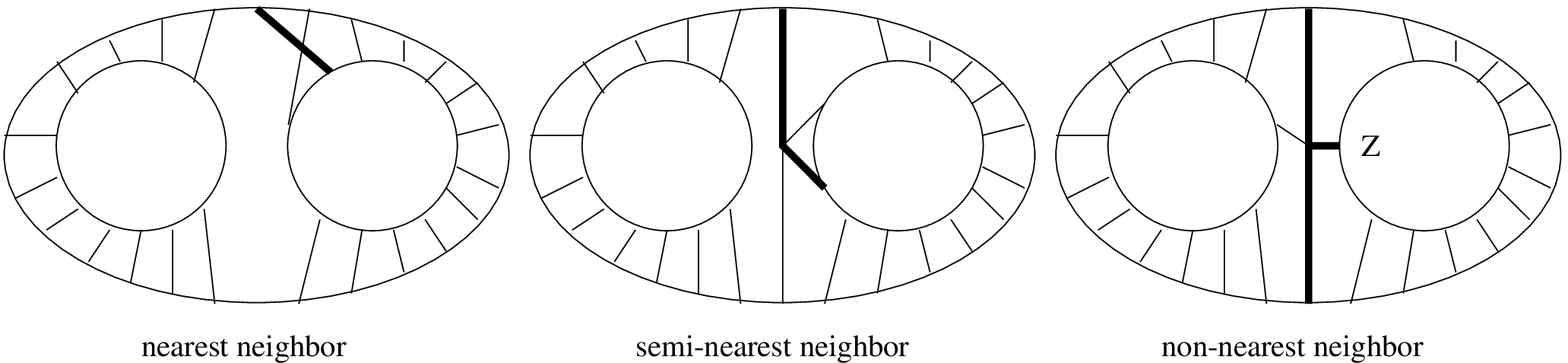}}

\noindent
$\bullet$ The nearest neighbor interaction\foot{This terminology was
first introduced in \ConstableHW.} vertex where two fields on each
operator coupled by the interaction 
sit next
to each other, involves one impurity in the $in$-operator,
   and the same impurity and $Z$ in the $out$-operator.
This interaction can occur at ${\cal O}(J)$ sites along the smaller
trace operator and we have 
to sum over the position of the interaction in the trace. 

\noindent
$\bullet$
The semi-nearest neighbor interaction vertex has two fields on one side
sitting
next to each other but the two fields  on the other side are separated by
${\cal O}(J)$
sites. The vertex can be inserted only at a particular place along the
trace and
so we do not sum over the position of the vertex.

\noindent
$\bullet$
The non-nearest neighbor interaction vertex has the two fields
on each side of the interaction point 
separated by ${\cal O}(J)$ sites. In this vertex, two
impurities or
$\bar{Z}$ are involved in the two operators  and this is possible only when
we have
two identical impurities in each operator. This interaction can also
occur at a specific location in the trace, so we do not sum over the
position of the vertex. 

The contribution of each interaction vertex is given as:
\eqn\IntNearest{ \eqalign{
I^{\rm nearest}_{n_i, p_i}(l_i)=\;&
{1\over \sqrt{JJ_1}}{g^2 N\over 8\pi^2}
(1-t_i)(1-\bar{s}_i) (t_i\bar{s}_i)^{l_i}
\qquad \hbox{for } i\in {\cal I}_1,\cr
&\quad {\rm or}\cr
&{1\over \sqrt{JJ_2}}{g^2 N\over 8\pi^2}(1-t_i)(1-\bar{s}_i)
t_i^{J_1}(t_i\bar{s}_i)^{l_i} \qquad \hbox{for } i\in {\cal I}_2,}}
\eqn\IntSemi{ \eqalign{
I^{\rm semi-nearest}_{n_i,p_i} =\;&
- {1\over \sqrt{JJ_1}}{g^2 N\over 8\pi^2}
[(1-t_i)+(1-\bar{s}_i)](1-t_i^{J_1}) \qquad \hbox{for }
i\in {\cal I}_1,\cr
&\quad {\rm or}\cr
&{1\over \sqrt{JJ_2}}{g^2 N\over 8\pi^2}
[(1-t_i)+(1-\bar{s}_i)](1-t_i^{-J_2})  \qquad \hbox{for }
i\in {\cal I}_2,}}
\eqn\IntNon{ \eqalign{
I^{\rm non-nearest}_{n_i,n_j,p_i,p_j} =\;&
-{1\over\sqrt{JJJ_1J_1}}
{g^2N\over 8\pi^2}(1-t_i^{J_1})(1-t_j^{J_1})
\qquad \hbox{for } i,j\in {\cal I}_1,\cr
&\quad {\rm or}\cr
&-{1\over\sqrt{JJJ_2J_2}}
{g^2N\over 8\pi^2}(1-t_i^{J_1})(1-t_j^{J_1})
\qquad \hbox{for } i,j\in {\cal I}_2,\cr
&\quad {\rm or}\cr
&{1\over\sqrt{JJJ_1J_2}}
{g^2N\over 8\pi^2}(1-t_i^{J_1})(1-t_j^{J_1})
\qquad \hbox{for } i\in {\cal I}_1,j\in{\cal I}_2,
}}
where $l_i$ in $I^{\rm nearest}_{n_i, p_i}(l_i)$ denotes
the position of the nearest neighbor interaction vertex to be
summed over. Here each factor of $1/\sqrt{J}$ or
   $1/\sqrt{J_r}\;\;(r=1,2)$ comes from each impurity participating in
the interaction.
The rest of impurities in the $in$-operator are freely contracted with
the remaining impurities  in the $out$-operator and each free contraction
contributes
\eqn\FreePhase{
{1\over\sqrt{JJ_1}}(t_i\bar{s}_i)^{l_i} \;\; {\rm for}\;\;i\in {\cal I}_1
\qquad{\rm or}\qquad
{1\over \sqrt{JJ_2}}t_i^{J_1}(t_i \bar{s}_i)^{l_1}
   \;\; {\rm for}\;\;i\in {\cal I}_2.}

Now we have to multiply all the the different contributions, coming
from the interaction vertex and the free contractions and sum over all possible
positions of the impurities.
However, the whole summation is simply
factorized in the large $J$ limit 
into sums over each contribution since each contribution 
is independent of the positions of the rest of impurities:
\ifig\Seminearest{The factorization property of gauge theory amplitudes.}
{\epsfxsize5in\epsfbox{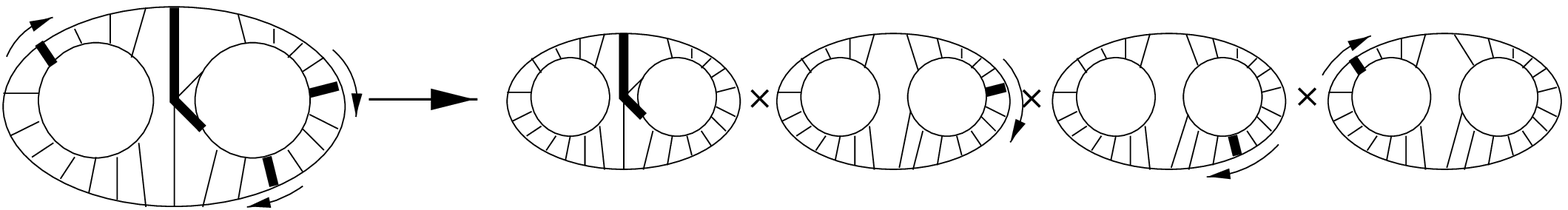}}
The computation of each contribution then essentially reduces to the one or
  two-impurity cases. This factorization property, which as we have
seen earlier has an analog in the string field theory computation,
will turn out to 
be useful in comparing the gauge theory and string theory expressions
for Feynman diagrams. As we will see, the effect of the prefactor
interaction  $(r,m)\;\hbox{---------------}
\!\!\!\!\!\!\!\!\!\!\!\!\!\!\!\!\!\!\times\;\;\;\;\;\;\;\,(s,l)$ in
string field theory is essentially captured by the interaction vertex
in gauge theory 
while the sum over free contractions in gauge theory capture the
Neumann matrices. 

Now, let us start to compute the string field theory 
 amplitudes and compare them
   with the gauge theory results. As discussed earlier, there are two
 cases to be considered.

{\it 1) Case 1 : Diagrams without self-contraction}

First, let us consider a particular way of contracting the oscillators
    without
self-contractions.
In this case, without loss of generality, we can assume
      that $d_i = e_i$ and take the
$d_i$-th oscillator to contract
with $e_i$-th oscillator for all $i\in \{1,\cdots,n\}$.
More specifically, the $j$-th oscillator in string 3 contracts with
the $j$-th oscillator in string 1
for $j\in {\cal I}_1$ and the $k$-th oscillator in string 3 contracts
with the
$k$-th oscillator in string 2 for
$k\in {\cal I}_2$. On the string field theory side, using the
Feynman rules in \Feynmm\ we can compute the
matrix elements between these states as in the previous section:
\eqn\ampli{\eqalign{
&{1\over \mu}\la (d_i,n_i)|H_3|(d_i,p_i);{\cal I}_1,{\cal I}_2;y\ra\ra
\cr
=&-(-1)^n{y(1-y)\over 2}\left\{\sum_{l\in {\cal
I}_1}\left[\left({\omega_{p_l(1)}\over \mu y}
-{\omega_{n_l(3)}\over
      \mu}\right){\tilde N}_{p_l,-n_l}^{(13)} \prod_{j\in {\cal 
I}_1-\{l\}}
{\tilde N}_{p_j,n_j}^{(13)}
      \prod_{k\in {\cal I}_2} {\tilde N}_{p_k,n_k}^{(23)}\right]
\right.\cr
&\qquad\left.+\sum_{l\in {\cal I}_2}\left[\left({\omega_{p_l(2)}\over
\mu(1-y)}-
{\omega_{n_l(3)}\over \mu}\right)
{\tilde N}_{p_l,-n_l}^{(23)}  \prod_{j\in {\cal I}_1} {\tilde
N}_{p_j,n_j}^{(13)}
      \prod_{k\in {\cal I}_2-\{l\}} {\tilde N}_{p_k,n_k}^{(23)}
\right]\right\}.}}
Now let us explain how to match each term above with
      specific Feynman diagrams in gauge theory.

$\bullet$  $l \in {\cal I}_1$
\ifig\Seminearest{Diagrams without
self-contractions, $l\in{\cal I}_1$. 
The numbers represent the direction of 
the SFT oscillators and the corresponding 
gauge theory impurities.}
{\epsfxsize4in\epsfbox{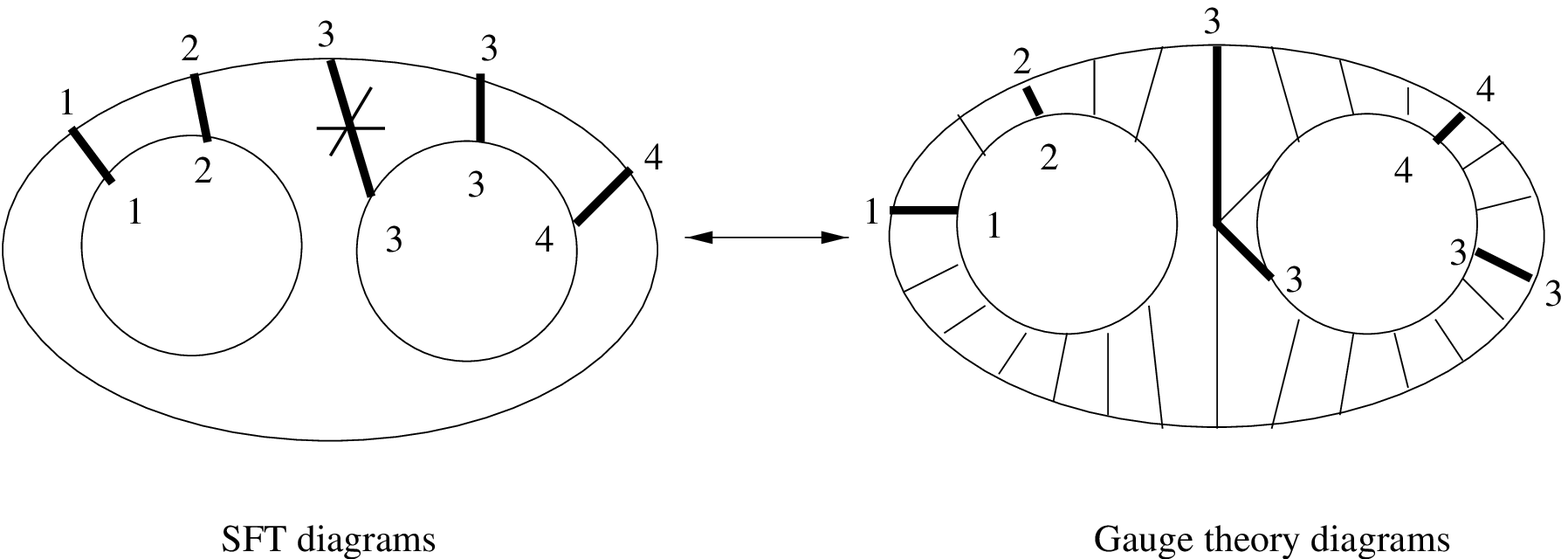}}

For each $l\in{\cal I}_1$, the particular term
\eqn\ltermONE{
(-1)^n{y(1-y)\over 2} \left({\omega_{n_l(3)}\over \mu}
-{\omega_{p_l(1)}\over \mu y}\right){\tilde N}_{p_l,-n_l}^{(13)}
\prod_{j\in{\cal I}_1-\{l\}}{\tilde N}_{p_j,n_j}^{(13)}
\prod_{k\in{\cal I}_2}{\tilde N}_{p_k,n_ k}^{(23)},}
arises when the $l$-th oscillator in string 1 and string 3 go
through the prefactor and contract while the rest of the oscillators
get contracted among themselves. The pair of $l$-th oscillators produce
\eqn\lterm{
{1\over 2}\left({\omega_{n_l(3)}\over\mu}
-{\omega_{p_l(1)}\over\mu y}\right){\tilde N}_{p_l,-n_l}^{(13)}
\simeq{1\over 4\mu^2}\left(n_l-{p_l\over y}\right)^2
{\tilde N}_{p_l,n_l}^{(13)},}
where we have used the large $\mu$ relation
\eqn\largemuNe{
{\tilde N}_{p,-n}^{(r3)}\simeq{n-{p\over y}\over n+{p\over y}}{\tilde
N}_{p,n}^{(r3)}\qquad r=1\ \hbox{or}\ 2.}
The other pairs of oscillators bring down one Neumann coefficient
${\tilde N}_{p_j,n_j}^{(13)}$ or ${\tilde N}_{p_k,n_k}^{(23)}$.
Therefore the contribution to the Hamiltonian matrix element due to
this diagram is\foot{After going to the unit norm basis.}:
\eqn\contrifinla{\eqalign{
&{1\over\mu}\la(d_i,n_i)|H_3|(d_i,p_i);{\cal I}_1,{\cal I}_2;y\ra\ra
\Bigr|_l\cr&\simeq(-1)^n{1\over 4\mu^2}
\sqrt{{y(1-y)\over J}}\left(n_l-{p_l\over y}\right)^2
\tN_{p_l,n_l}^{(13)}
\prod_{j\in {\cal I}_1-\{l\}}\tN_{p_j,n_j}^{(13)}
\prod_{k\in {\cal I}_2}\tN_{p_k,n_ k}^{(23)}.}}

We claim that this particular term corresponds to the interaction
Feynman diagrams where two $\phi_l$'s are involved in the interaction
vertex and the rest of the impurities are freely contracted. The
contributions come from two classes of diagrams. The nearest neighbor
diagrams give
\eqn\lnearest{\eqalign{
\Gamma^{(1)}_{\{n_i\},\{p_i;{\cal I}_1,{\cal I}_2\}y}
\Bigr|_l^{\rm nearest}
&=\sqrt{y(1-y)\over J}\times\biggl[
{g^2 N\over 8\pi^2}(1-t_l)(1-\bar{s}_l) {1\over \sqrt{JJ_1}}
\sum_{a=0}^{J_1-1} (t_l\bar{s}_l)^{a}\biggr]\cr
&\quad\times\prod_{j\in {\cal I}_1-\{l\}}
{1\over \sqrt{JJ_1}} \sum_{a=0}^{J_1-1} (t_j\bar{s}_j)^{a}
\prod_{k\in {\cal I}_2}{1\over \sqrt{JJ_2}}\sum_{a=0}^{J_2-1}
t_k^{J_1}(t_k\bar{s}_k)^{a},
}}
whereas the semi-nearest neighbor diagrams contribute
\eqn\lsemi{ \eqalign{
\Gamma^{(1)}_{\{n_i\},\{p_i;{\cal I}_1,{\cal I}_2\}y}
\Bigr|_l^{\rm semi-nearest}
&=\sqrt{y(1-y)\over J}\times\biggl[-{1\over\sqrt{JJ_1}}
{g^2N\over 8\pi^2}[(1-t_l)+(1-\bar{s}_l)](1-t_l^{J_1})\biggr]\cr
&\quad\times \prod_{j\in {\cal I}_1-\{l\}}{1\over \sqrt{JJ_1}}
\sum_{a=0}^{J_1-1}
(t_j\bar{s}_j)^{a}
\prod_{k\in {\cal I}_2}{1\over \sqrt{JJ_2}}\sum_{a=0}^{J_2-1}
t_k^{J_1}(t_k\bar{s}_k)^{a},}}
where the phases are defined as $t_i=\exp(2\pi in_i/J)$, and
$s_j=\exp(2\pi ip_j/J_1)$ for $j\in{\cal I}_1$ and
$s_k=\exp(2\pi ip_k/J_2)$ for $k\in{\cal I}_2$.
The subscript $l$ means that only Feynman diagrams with $\phi_l$'s
involved in the interaction vertex are included. The first factor in
\lnearest\ and \lsemi\ comes from the interaction vertices involving
$\phi_l$ and the rest of the expression comes from free contraction of
the other impurities. We can compute each factor and express it in
terms of purely string field theory quantities and show that the
interaction essentially
captures the energy difference factor in the string theory
computation while the free contractions yield the Neumann matrices.
For $j \in {\cal I}_1$ and $k \in {\cal I}_2$, the free contraction
contribution is:
\eqn\phasesum{\eqalign{{1\over\sqrt{JJ_1}}\sum_{a=0}^{J_1-1}
(t_j\bar{s}_j)^{a}&\simeq(-1)^{n_j+p_j+1}e^{i\pi n_j y}
\tilde{N}^{(13)}_{p_j,n_j},\cr
{1\over \sqrt{JJ_2}}\sum_{a=0}^{J_2-1}
t_k^{J_1}(t_k \bar{s}_k)^a
&\simeq(-1)^{n_k+1}e^{i\pi n_k y} \tilde{N}^{(23)}_{p_k,n_k},}}
while the interaction vertex contribution is:
\eqn\interactionphase{\eqalign{
{1\over\sqrt{JJ_1}}{g^2N\over 8\pi^2}(1-t_l)(1-\bar{s}_l)
\sum_{a=0}^{J_1-1}(t_l\bar{s}_l)^{a}
&\simeq (-1)^{n_l+p_l+1}e^{i\pi n_l y}\times
{\lambda^\prime\over 2}\left({n_lp_l\over y}\right)
\tilde{N}^{(13)}_{p_l,n_l},\cr
-{1\over\sqrt{JJ_1}}{g^2 N\over 8\pi^2}
[(1-t_l)+(1-\bar{s}_l)](1-t_l^{J_1})
&\simeq (-1)^{n_l+p_l+1}e^{i\pi n_l y}\times{\lambda^\prime\over 2}
\left(n_l-{p_l\over y}\right)^2\tilde{N}^{(13)}_{p_l,n_l}.}}
Altogether, we obtain
\eqn\lGammaONE{\eqalign{
&\Gamma^{(1)}_{\{n_i\},\{p_i;{\cal I}_1,{\cal I}_2\}y}\Bigr|_l\cr
&\simeq (-1)^n{\lambda^\prime\over 2}\sqrt{y(1-y)\over J}
\left[\left(n_l-{p_l\over y}\right)^2+n_l{p_l\over y}\right]
\tilde{N}^{(13)}_{p_l,n_l}
\prod_{j\in {\cal I}_1-\{l\}}\tilde{N}^{(13)}_{p_j,n_j}
\prod_{k\in {\cal I}_2}\tilde{N}^{(23)}_{p_k,n_k}.}}
Notice that all the phase factors except $(-1)^n$ disappear upon imposing
the level-matching conditions.
In order to compare with the string theory result, we must evaluate these
expressions in the string field theory basis \anombasis .
In order to compute,
\eqn\lTildeGamma{
\tilde\Gamma^{(1)}\bigr|_l=\Gamma^{(1)}\bigr|_l-{1\over 2}\{G^{(1)},
\Gamma^{(0)}\bigr|_l\},}
we also need to compute $G^{(1)}$ and $\Gamma^{(0)}\bigr|_l$. They are
given by\foot{Here we use the large $\mu$ relation \phasesum\ to
rewrite $G^{(1)}$ in terms of string field theory quantities.}:
\eqn\GOnelGammaZero{\eqalign{
G^{(1)}_{\{n_i\},\{p_i;{\cal I}_1,{\cal I}_2\}y}
&=\sqrt{y(1-y)\over J}
\prod_{j\in{\cal I}_1}{1\over\sqrt{JJ_1}}
\sum_{a=0}^{J_1-1}(t_j\bar{s}_j)^{a}
\prod_{k\in{\cal I}_2}{1\over\sqrt{JJ_2}}
\sum_{a=0}^{J_2-1}t_k^{J_1}(t_k\bar{s}_k)^{a}\cr
&\simeq (-1)^n\sqrt{y(1-y)\over J}
\prod_{j\in {\cal I}_1}\tilde{N}^{(13)}_{p_j,n_j}
\prod_{k\in {\cal I}_2}\tilde{N}^{(23)}_{p_k,n_k},\cr
\Gamma^{(0)}_{\{n_i\},\{m_i\}}\bigr|_l
&={\lambda^\prime\over 2}n_l^2\prod_{i}\delta_{n_i,m_i},\cr
\Gamma^{(0)}_{\{p_i;{\cal I}_1,{\cal I}_2\}y,
\{q_i;{\cal I}_1,{\cal I}_2\}z}\bigr|_l
&={\lambda^\prime\over 2}\biggl({p_l\over y}\biggr)^2
\delta_{y,z}\prod_{i}\delta_{p_i,q_i}.}}
Hence,
\eqn\lTildeGammaOne{
\tilde\Gamma^{(1)}_{\{n_i\},\{p_i;{\cal I}_1,{\cal I}_2\}y}\Bigr|_l
\simeq (-1)^n{\lambda^\prime\over 4}\sqrt{y(1-y)\over J}
\left(n_l-{p_l\over y}\right)^2\tilde{N}^{(13)}_{p_l,n_l}
\prod_{j\in {\cal I}_1-\{l\}}\tilde{N}^{(13)}_{p_j,n_j}
\prod_{k\in {\cal I}_2}\tilde{N}^{(23)}_{p_k,n_k}.}
which precisely reproduces the string field theory result \contrifinla .

$\bullet$  $l\in{\cal I}_2$
\ifig\Seminearest{Diagrams without self-contractions,
$l\in {\cal I}_2$.}{\epsfxsize4in\epsfbox{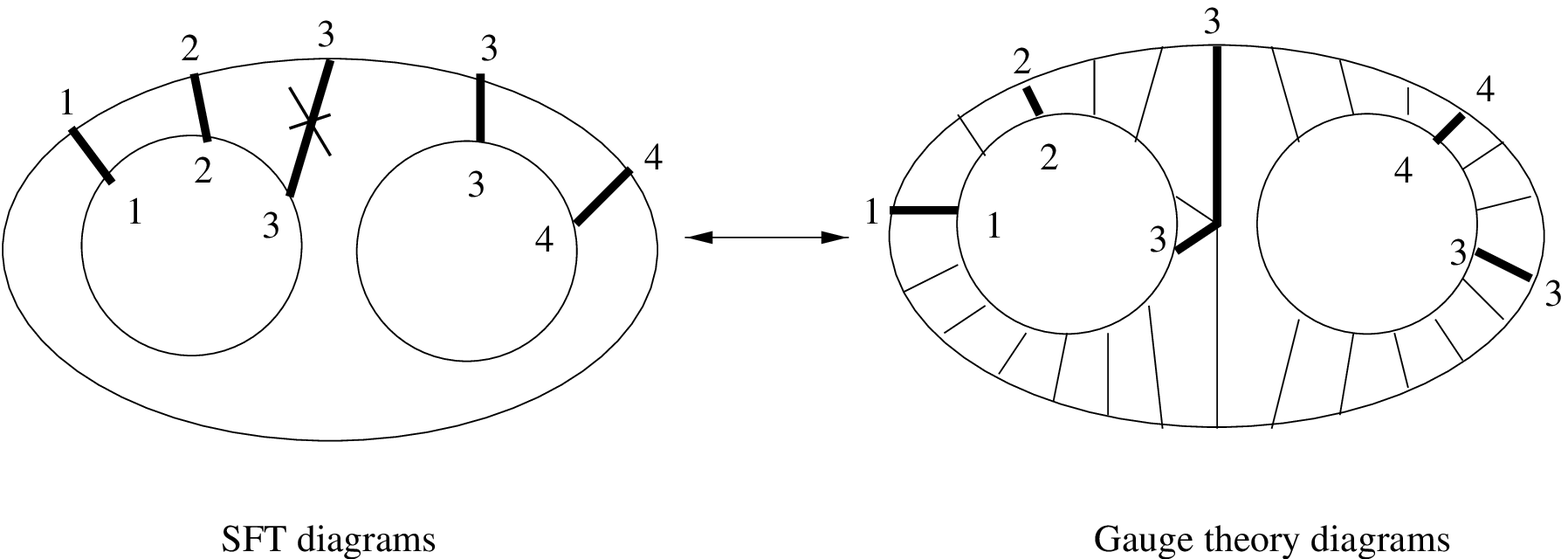}}

Now we consider the other type of contraction in the string field
theory computation, where the prefactor acts on the $l$-th oscillator
in string 2 and string 3. The expression for this diagram is:
\eqn\ltermTWO{ \eqalign{
&(-1)^n{y(1-y)\over 2} \left({\omega_{n_l(3)}\over \mu}
-{\omega_{p_l(2)}\over \mu(1- y)}\right){\tilde N}_{p_l,-n_l}^{(23)}
\prod_{j\in {\cal I}_1} {\tilde N}_{p_j,n_j}^{(13)} \prod_{k\in {\cal
I}_2-\{l\}}
{\tilde N}_{p_k,n_k}^{(23)}\cr}.}
As before, it is convenient to express the contribution from the
prefactor as:
\eqn\ltermTWO{{1\over 2}
\left({\omega_{n_l(3)}\over\mu}
-{\omega_{p_l(2)}\over\mu(1-y)}\right)
{\tilde N}_{p_l,-n_l}^{(23)} \simeq
{1\over 4\mu^2} \left( n_l - {p_l \over 1- y}\right)^2
{\tilde N}_{p_l,n_l}^{(23)},}
where we have used the large $\mu$ relation \largemuNe. Therefore, the
contribution of this diagram to the Hamiltonian matrix element of
unit normalized states is:
\eqn\contrifinlab{\eqalign{
&{1\over\mu}\la(d_i,n_i)|H_3|(d_i,p_i);{\cal I}_1,{\cal I}_2;
y\ra\ra\Bigr|_l\cr
&\simeq(-1)^n{1\over 4\mu^2}
\sqrt{{y(1-y)\over J}}\left(n_l-{p_l\over 1-y}\right)^2
{\tilde N}_{p_l,n_l}^{(23)}
\prod_{j\in {\cal I}_1}{\tilde N}_{p_j,n_j}^{(13)}
\prod_{k\in {\cal I}_2-\{l\}}{\tilde N}_{p_k,n_ k}^{(23)}.}}

The corresponding gauge theory diagrams are again classified into two
classes.
The nearest neighbor diagrams yields
\eqn\lnearestTWO{\eqalign{
\Gamma^{(1)}_{\{n_i\},\{p_i;{\cal I}_1,{\cal I}_2\}y}
\Bigr|_l^{\rm nearest}
&=\sqrt{y(1-y)\over J}\times
\biggl[{g^2N\over 8\pi^2}(1-t_l)(1-\bar{s}_l) {1\over \sqrt{JJ_2}}
\sum_{b=0}^{J_2-1} t_l^{J_1}(t_l\bar{s}_l)^{b}\biggr]\cr
&\quad\times\prod_{j\in {\cal I}_1}
{1\over \sqrt{JJ_1}} \sum_{a=0}^{J_1-1} (t_j\bar{s}_j)^{a}
\prod_{k\in {\cal I}_2-\{l\}}{1\over \sqrt{JJ_2}}\sum_{b=0}^{J_2-1}
t_k^{J_1}(t_k\bar{s}_k)^{b},}}
whereas the semi-nearest neighbor diagrams contribute
\eqn\lsemi{ \eqalign{
\Gamma^{(1)}_{\{n_i\},\{p_i;{\cal I}_1,{\cal I}_2\}y}
\Bigr|_l^{\rm semi-nearest}
&=\sqrt{y(1-y)\over J}\times\biggl[{1\over\sqrt{JJ_2}}
{g^2 N\over 8\pi^2}[(1-t_l)+(1-\bar{s}_l)](1-t_l^{J_1})\biggr]\cr
&\quad\times\prod_{j\in{\cal I}_1}{1\over\sqrt{JJ_1}}
\sum_{a=0}^{J_1-1}(t_j\bar{s}_j)^{a}
\prod_{k\in {\cal I}_2-\{l\}}{1\over \sqrt{JJ_2}}\sum_{b=0}^{J_2-1}
      t_k^{J_1}(t_k\bar{s}_k)^{b}.}}
We can also express the various contributions in terms of string
field theory quantities.
The interaction vertex contribution is given by
\eqn\interactionphase{\eqalign{
{g^2 N\over 8\pi^2}(1-t_l)(1-\bar{s}_l) {1\over \sqrt{JJ_2}}
\sum_{b=0}^{J_2-1} t_l^{J_1}(t_l\bar{s}_l)^{b}
&\simeq (-1)^{n_l+1}e^{i\pi n_l y}\times
{\lambda^\prime \over 2} \left({n_lp_l\over 1- y}\right)
\tilde{N}^{(23)}_{p_l,n_l},\cr
{1\over \sqrt{JJ_2}} {g^2 N\over
8\pi^2}[(1-t_l)+(1-\bar{s}_l)](1-t_l^{J_1})
&\simeq (-1)^{n_l+1}e^{i\pi n_l y}\times {\lambda^\prime \over 2}
\left(n_l-{p_l\over 1-y}\right)^2\tilde{N}^{(23)}_{p_l,n_l},}}
whereas the free contraction \phasesum\ yields the product of
Neumann matrix after imposing the level matching constraint.

In order to compute the matrix of anomalous dimensions in the string
field theory basis we need also $G^{(1)}$ and
$\Gamma^{(0)}\Big|_l$. It is easy to show that these quantities are
the same as in \GOnelGammaZero\ except for the last formula which can
be correctly obtained by replacing $y\rightarrow 1-y$. Therefore,
using \lTildeGamma, we obtain:
\eqn\lTildeGammaOne{
\tilde\Gamma^{(1)}_{\{n_i\},\{p_i;{\cal I}_1,{\cal I}_2\}y}
\Big|_l \simeq
(-1)^n{\lambda^\prime \over 4} \sqrt{y(1-y)\over J}
\left(n_l-{p_l\over 1-y}\right)^2 \tilde{N}^{(23)}_{p_l,n_l}
\prod_{j\in {\cal I}_1}  \tilde{N}^{(13)}_{p_j,n_j}
\prod_{k\in {\cal I}_2-\{l\}}  \tilde{N}^{(23)}_{p_k,n_k},}
and again we find agreement with the string theory result \contrifinlab .

{\it 2) Terms with self-contractions}

As explained in the beginning of this section, to leading order in the
$1/ \mu$ expansion we can have at most one self-contraction in
string 3 and the prefactor has to go through any of the oscillators
involved in the self-contraction.

Without loss of generality, we can assume that
$d_1=d_2$, $e_1=e_2$, $d_i=e_i$ for $i \in \{3,\cdots,n\}$ and we will
consider contractions between $d_1-d_2$, $e_1-e_2$, and $d_i-e_i$
for $i \in \{3,\cdots,n\}$. There are three cases depending on
how the 1st and the 2nd impurities are distributed on the two-string
state and the double-trace operator: $1,2\in {\cal I}_1$,
$1,2\in {\cal I}_2$ and $1\in {\cal I}_1$, $2\in {\cal I}_2$.

$\bullet$ $1,2\in {\cal I}_1$
\ifig\Seminearest{Diagrams with self-contractions, $1,2\in {\cal I}_1$ .}
{\epsfxsize4in\epsfbox{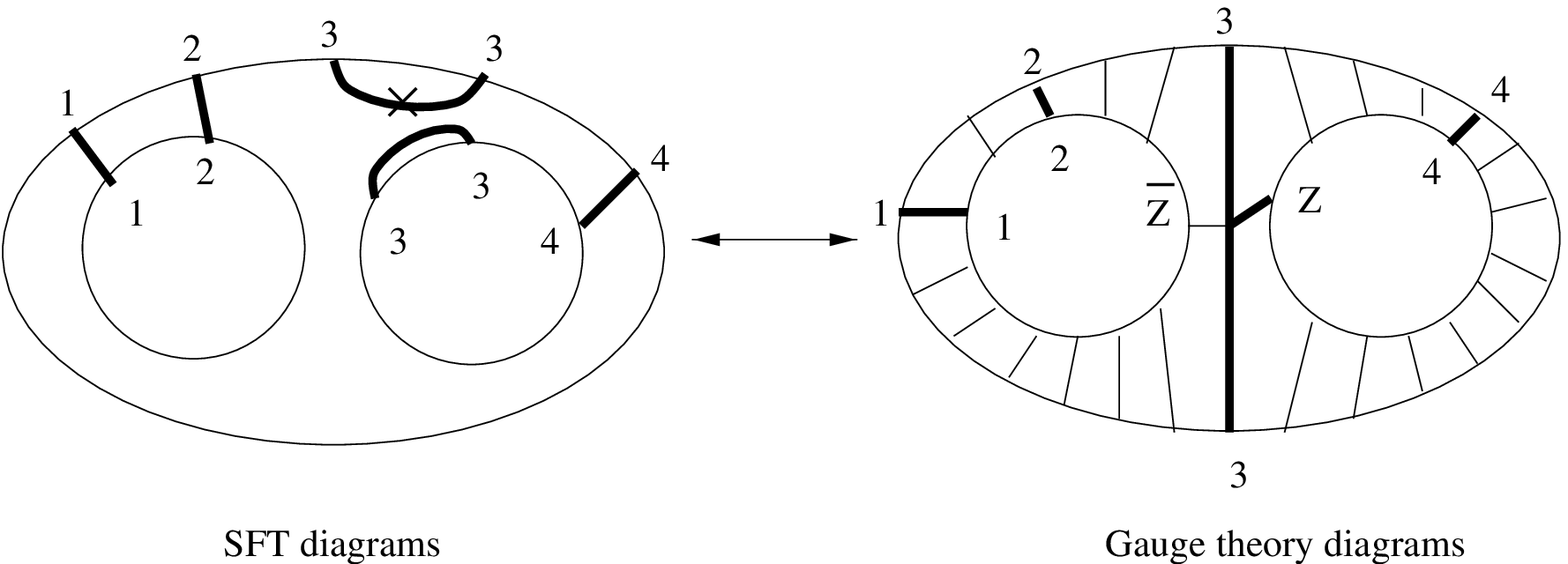}}

The string theory computation of this particular Feynman diagram is:
\eqn\SFTampli{\eqalign{
&{1\over\mu}\la (d_i,n_i)|H_3
|(e_i,p_i);{\cal I}_1,{\cal I}_2;y\ra\ra\cr
&=-(-1)^n{y(1-y)\over 2}\left[
\left({\omega_{p_1(1)}\!+\!\omega_{p_2(1)}\over\mu y}\right)
\tilde{N}_{n_1,n_2}^{(33)}\tilde{N}_{p_1,-p_2}^{(11)}-
\left({\omega_{n_1(3)}\!+\!\omega_{n_2(3)}\over\mu}\right)
\tilde{N}_{n_1,-n_2}^{(33)}\tilde{N}_{p_1,p_2}^{(11)}\right]\cr
&\quad\times\prod_{j\in{\cal I}_1-\{1,2\}}\tilde{N}_{p_j,n_j}^{(13)}
\prod_{k\in{\cal I}_2}\tilde{N}_{p_k,n_k}^{(23)}.}}
The first line in \SFTampli\ is due to the self-contractions while the
rest is due to the contraction between oscillators in the single
string state with the two-string state.

The self-contraction contribution is to the leading order in $1/\mu$:
\eqn\selfint{\eqalign{
&-{1\over2}
\left[\left({\omega_{p_1(1)}+\omega_{p_2(1)}\over\mu y}\right)
\tilde{N}_{n_1,n_2}^{(33)}\tilde{N}_{p_1,-p_2}^{(11)}-
\left({\omega_{n_1(3)}+\omega_{n_2(3)}\over \mu
}\right)\tilde{N}_{n_1,-n_2}^{(33)}\tilde{N}_{p_1,p_2}^{(11)} \right]\cr
\simeq& -2\tilde{N}_{n_1,n_2}^{(33)}\tilde{N}_{p_1,p_2}^{(11)} .}}
Therefore, the matrix element of unit normalized states is given by
\eqn\SFTamplifin{\eqalign{
&{1\over\mu}\la(d_i,n_i)|H_3
|(e_i,p_i);{\cal I}_1,{\cal I}_2;y\ra\ra\Big|_{1-2;1-2}\cr
&\simeq-2(-1)^n{\sqrt{{y(1-y)\over J}}}
\tilde{N}_{n_1,n_2}^{(33)}\tilde{N}_{p_1,p_2}^{(11)}
\prod_{j\in{\cal I}_1-\{1,2\}}\tilde{N}_{p_j,n_j}^{(13)}
\prod_{k\in {\cal I}_2}\tilde{N}_{p_k,n_k}^{(23)}.}}

We now show that the corresponding gauge theory diagrams are those
with an interaction vertex involving
     $\phi_{d_1},\phi_{d_2}$ or $\bar{Z}$ in
${\cal O}^J_{(d_i,n_i)}$ and $\phi_{e_1},\phi_{e_2}$ or $\bar{Z}$
in ${\cal T}^{J,y}_{(e_i,p_i);{\cal I}_1,{\cal I}_2}$.
In this case,  only non-nearest interaction diagrams
contribute and the result is:
\eqn\selfsemi{ \eqalign{
\Gamma^{(1)}_{\{n_i\},\{p_i;{\cal I}_1,{\cal I}_2\}y}
\Big|_{1-2;1-2}^{\rm non-nearest}
&={\sqrt{y(1-y)\over J}}\times\biggl[-{1\over\sqrt{JJJ_1J_1}}
{g^2N\over 8\pi^2}(1-t_1^{J_1})(1-t_2^{J_1})\biggr]\cr
&\quad\times\prod_{j\in {\cal I}_1-\{1,2\}}{1\over\sqrt{JJ_1}}
\sum_{a=0}^{J_1-1}(t_j\bar{s}_j)^{a}
\prod_{k\in {\cal I}_2}{1\over \sqrt{JJ_2}}\sum_{b=0}^{J_2-1}
t_k^{J_1}(t_k\bar{s}_k)^{b}.}}
The interaction contribution reduces in the BMN limit to
\eqn\selffirstfactor{\eqalign{
-{1\over\sqrt{JJJ_1J_1}}{g^2N\over 8\pi^2}(1-t_1^{J_1})(1-t_2^{J_1})
&=e^{\pi i(n_1+n_2)y}\lambda^\prime
{\sin(\pi n_1y)\sin(\pi n_2y)\over 2\pi^2y}\cr
&\simeq-2(-1)^{n_1+n_2+p_1+p_2}e^{\pi i(n_1+n_2)y}
\tilde{N}_{n_1,n_2}^{(33)}\tilde{N}_{p_1,p_2}^{(11)} ,}}
while the rest can be rewritten in string field theory language using
\phasesum . Again the various phase factors disappear after imposing
     the level matching condition on each trace.

In order to compare with string field theory we must go to the string
field theory basis. However, the particular class of Feynman
diagrams we are considering, which are those with an interaction vertex
    involving
     $\phi_{d_1},\phi_{d_2}$ or $\bar{Z}$ in
${\cal O}^J_{(d_i,n_i)}$ and $\phi_{e_1},\phi_{e_2}$ or $\bar{Z}$
in ${\cal T}^{J,y}_{(e_i,p_i);{\cal I}_1,{\cal I}_2}$ do not
contribute to $\Gamma^{(0)}\Big|_l$. Therefore, in this case
\lTildeGamma\ yields:
\eqn\selfGammaONE{
{\tilde\Gamma}^{(1)}_{\{n_i\},\{p_i;{\cal I}_1,{\cal I}_2\}y}
\Big|_{1-2;1-2}^{\rm non-nearest}
\simeq -2(-1)^n {\sqrt{y(1-y) \over J}}
\tilde{N}_{n_1,n_2}^{(33)}\tilde{N}_{p_1,p_2}^{(11)}
\prod_{j\in {\cal I}_1-\{1,2\}}\!\!\!\!\!\!\!\tilde{N}_{p_j,n_j}^{(13)}
\prod_{k\in {\cal I}_2}\tilde{N}_{p_k,n_k}^{(23)},}
which agrees with the SFT result \SFTampli.

\vfill\eject

$\bullet$ $1,2\in {\cal I}_2$
\ifig\Seminearest{Diagrams with self-contractions, $1,2\in {\cal I}_2$ .}
{\epsfxsize4in\epsfbox{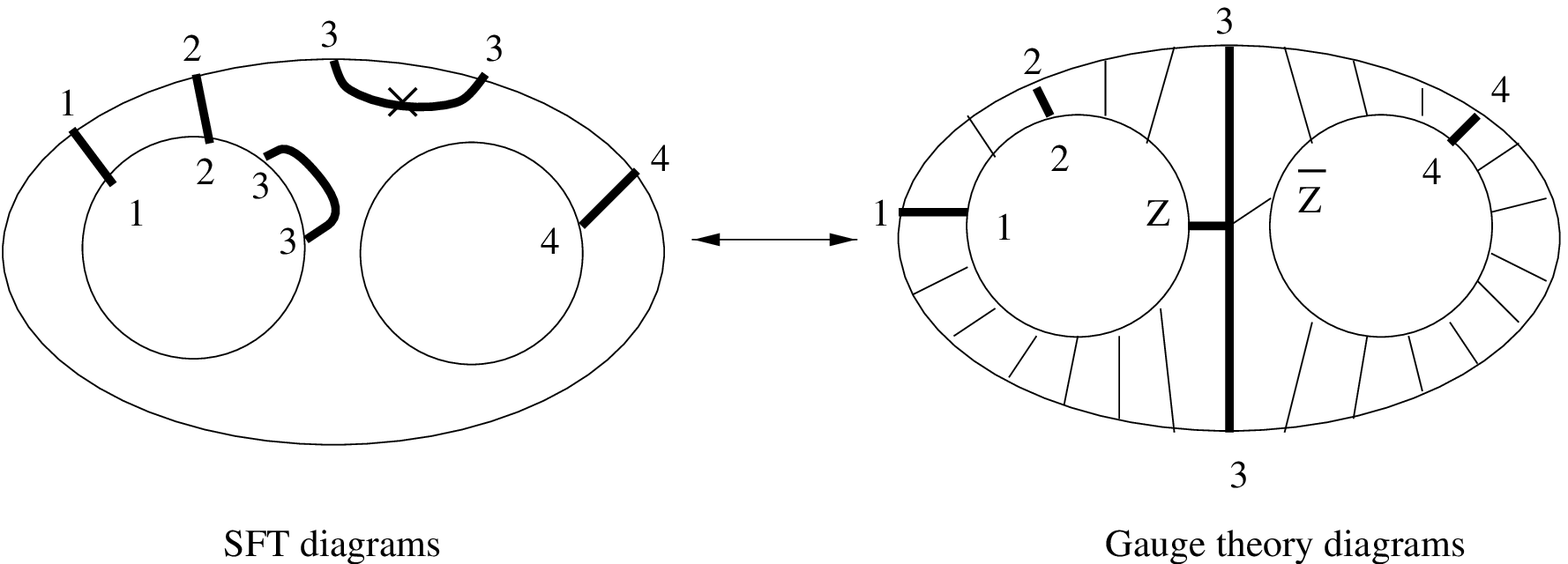}}

The string theory computation of this particular Feynman diagram is
similar to the previous one:
\eqn\SFTamplib{\eqalign{
&{1\over\mu}\la (d_i,n_i)|H_3
|(e_i,p_i);{\cal I}_1,{\cal I}_2;y\ra\ra\cr
&=-(-1)^n{y(1-y)\over 2}\left[
\left({\omega_{p_1(2)}\!+\!\omega_{p_2(2)}\over\mu (1-y)}\right)
\tilde{N}_{n_1,n_2}^{(33)}\tilde{N}_{p_1,-p_2}^{(22)}-
\left({\omega_{n_1(3)}\!+\!\omega_{n_2(3)}\over\mu}\right)
\tilde{N}_{n_1,-n_2}^{(33)}\tilde{N}_{p_1,p_2}^{(22)}\right]\cr
&\quad\times\prod_{j\in{\cal I}_1}\tilde{N}_{p_j,n_j}^{(13)}
\prod_{k\in{\cal I}_2-\{1,2\}}\tilde{N}_{p_k,n_k}^{(23)}.}}

The self-contraction contribution is to the leading order in $1/\mu$:
\eqn\selfintb{\eqalign{
& -{1\over2}
\left[\left({\omega_{p_1(2)}+\omega_{p_2(2)}\over\mu (1-y)}\right)
\tilde{N}_{n_1,n_2}^{(33)}\tilde{N}_{p_1,-p_2}^{(22)}-
\left({\omega_{n_1(3)}+\omega_{n_2(3)}\over \mu
}\right)\tilde{N}_{n_1,-n_2}^{(33)}\tilde{N}_{p_1,p_2}^{(22)} \right]\cr
\simeq& -2\tilde{N}_{n_1,n_2}^{(33)}\tilde{N}_{p_1,p_2}^{(22)}.}}
Therefore, the matrix element of unit normalized states is given by
\eqn\SFTamplifinb{\eqalign{
&{1\over\mu}\la(d_i,n_i)|H_3
|(e_i,p_i);{\cal I}_1,{\cal I}_2;y\ra\ra\Big|_{1-2;1-2}\cr
&\simeq-2(-1)^n{\sqrt{{y(1-y)\over J}}}
\tilde{N}_{n_1,n_2}^{(33)}\tilde{N}_{p_1,p_2}^{(22)}
\prod_{j\in{\cal I}_1}\tilde{N}_{p_j,n_j}^{(13)}
\prod_{k\in {\cal I}_2-\{1,2\}}\tilde{N}_{p_k,n_k}^{(23)}.}}

We now show that the corresponding gauge theory diagrams are those
with an interaction vertex involving
     $\phi_{d_1},\phi_{d_2}$ or $\bar{Z}$ in
${\cal O}^J_{(d_i,n_i)}$ and $\phi_{e_1},\phi_{e_2}$ or $\bar{Z}$
in ${\cal T}^{J,y}_{(e_i,p_i);{\cal I}_1,{\cal I}_2}$.
In this case,  only non-nearest interaction diagrams
contribute and the result is:
\eqn\selfsemib{ \eqalign{
\Gamma^{(1)}_{\{n_i\},\{p_i;{\cal I}_1,{\cal I}_2\}y}
\Big|_{1-2;1-2}^{\rm non-nearest}
&={\sqrt{y(1-y)\over J}}\times\biggl[-{1\over\sqrt{JJJ_2J_2}}
{g^2N\over 8\pi^2}(1-t_1^{J_1})(1-t_2^{J_1})\biggr]\cr
&\quad\times\prod_{j\in {\cal I}_1}{1\over\sqrt{JJ_1}}
\sum_{a=0}^{J_1-1}(t_j\bar{s}_j)^{a}
\prod_{k\in {\cal I}_2-\{1,2\}}{1\over \sqrt{JJ_2}}\sum_{b=0}^{J_2-1}
t_k^{J_1}(t_k\bar{s}_k)^{b}.}}
The interaction contribution reduces in the BMN limit to
\eqn\selffirstfactorb{\eqalign{
-{1\over\sqrt{JJJ_2J_2}}{g^2N\over 8\pi^2}(1-t_1^{J_1})(1-t_2^{J_1})
&=e^{\pi i(n_1+n_2)y}\lambda^\prime
{\sin(\pi n_1y)\sin(\pi n_2y)\over 2\pi^2(1-y)}\cr
&\simeq-2(-1)^{n_1+n_2}e^{\pi i(n_1+n_2)y}
\tilde{N}_{n_1,n_2}^{(33)}\tilde{N}_{p_1,p_2}^{(22)}
,}}
while the rest can be rewritten in string field theory language using
\phasesum . Again the various phase factors disappear after imposing
     the level matching condition on each trace.

In order to compare with string field theory we must go to the string
field theory basis. However, the particular class of Feynman
diagrams we are considering, which are those with an interaction vertex
    involving
     $\phi_{d_1},\phi_{d_2}$ or $\bar{Z}$ in
${\cal O}^J_{(d_i,n_i)}$ and $\phi_{e_1},\phi_{e_2}$ or $\bar{Z}$
in ${\cal T}^{J,y}_{(e_i,p_i);{\cal I}_1,{\cal I}_2}$ do not
contribute to $\Gamma^{(0)}\Big|_l$. Therefore, in this case
\lTildeGamma\ yields:
\eqn\selfGammaONEb{
{\tilde\Gamma}^{(1)}_{\{n_i\},\{p_i;{\cal I}_1,{\cal I}_2\}y}
\Big|_{1-2;1-2}^{\rm non-nearest}
\simeq -2(-1)^n {\sqrt{y(1-y) \over J}}
\tilde{N}_{n_1,n_2}^{(33)}\tilde{N}_{p_1,p_2}^{(22)}
\prod_{j\in {\cal I}_1}\tilde{N}_{p_j,n_j}^{(13)}
\prod_{k\in {\cal I}_2-\{1,2\}}\!\!\!\!\!\!\! \tilde{N}_{p_k,n_k}^{(23)},}
which agrees with the SFT result \SFTamplib.

$\bullet$ $1\in {\cal I}_1$, $2\in {\cal I}_2$
\ifig\Seminearest{Diagrams with self-contractions, $1\in{\cal I}_1$,
$2\in{\cal I}_2$.}
{\epsfxsize4in\epsfbox{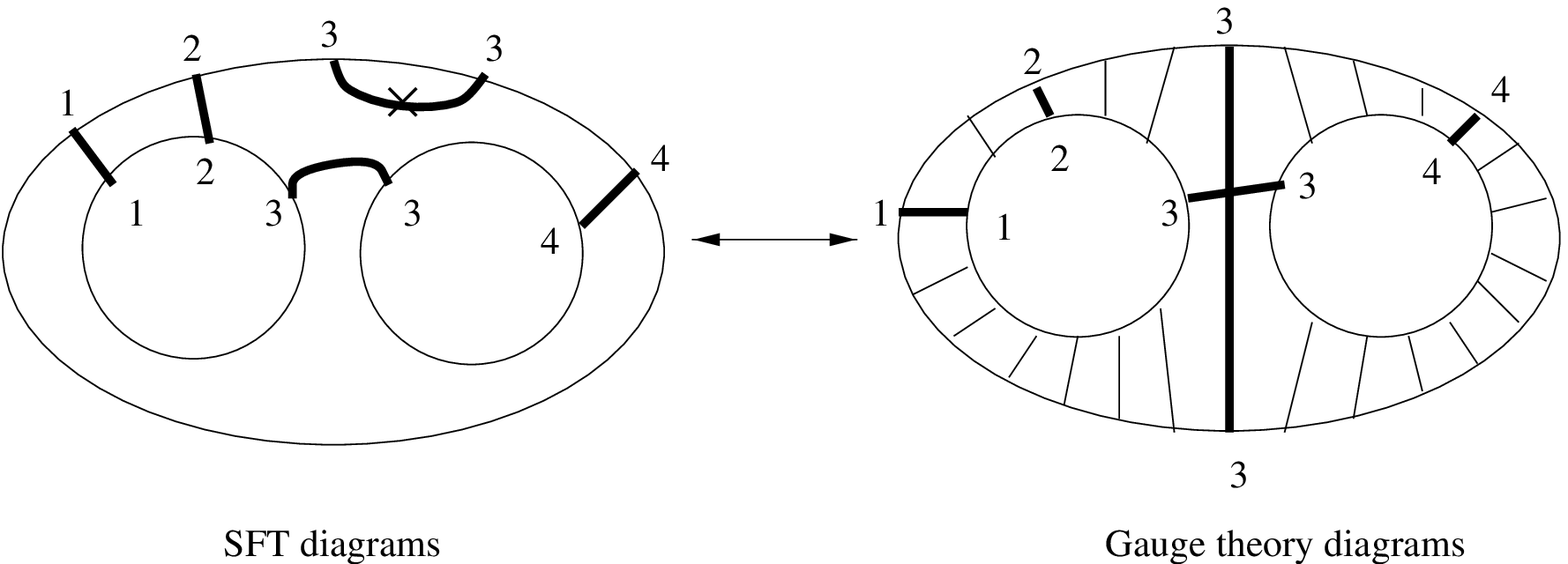}}

The string field theory  computation of this particular contraction
term is:
\eqn\SFTampliDiff{\eqalign{
&{1\over\mu}\la(d_i,n_i)|H_3
|(e_i,p_i);{\cal I}_1,{\cal I}_2;y\ra\ra\cr
&=-(-1)^n{y(1-y)\over 2}\!\left[\left(\!{\omega_{p_1(1)}\over\mu y}
\!+\!{\omega_{p_2(2)}\over\mu(1-y)}\!\right)\!
\tilde{N}_{n_1,n_2}^{(33)}\tilde{N}_{p_1,-p_2}^{(12)}
\!-\!\left(\!{\omega_{n_1(3)}\!+\!\omega_{n_2(3)}\over\mu}\!\right)\!
\tilde{N}_{n_1,-n_2}^{(33)}\tilde{N}_{p_1,p_2}^{(12)}\right]\cr
&\quad\times\prod_{j\in {\cal I}_1-\{1\}}\tilde{N}_{p_j,n_j}^{(13)}
\prod_{k\in{\cal I}_2-\{2\}}\tilde{N}_{p_k,n_k}^{(23)}.}}
The first factor which is the
result of the self-contraction between  the single and two-string state,
is to the leading order in $1/\mu$:
\eqn\selfintDiff{\eqalign{
\!\!& -{1\over2}\left[\!
\left(\!{\omega_{p_1(1)}\over\mu y}
+{\omega_{p_2(2)}\over\mu (1-y)}\!\right)
\tilde{N}_{n_1,n_2}^{(33)}\tilde{N}_{p_1,-p_2}^{(12)}
-\left(\!{\omega_{n_1(3)}+\omega_{n_2(3)}\over\mu}\!\right)
\tilde{N}_{n_1,-n_2}^{(33)}\tilde{N}_{p_1,p_2}^{(12)}\!\right]\cr
\simeq& -2\tilde{N}_{n_1,n_2}^{(33)}\tilde{N}_{p_1,p_2}^{(12)}.}}
Therefore, the contribution of this Feynman diagram to
the matrix element of unit normalized states is given by:
\eqn\SFTamplifinb{\eqalign{
&{1\over\mu}\la(d_i,n_i)|H_3
|(e_i,p_i);{\cal I}_1,{\cal I}_2;y\ra\ra\Big|_{1-2;1-2}\cr
&\simeq-2(-1)^n{\sqrt{{y(1-y)\over J}}}
\tilde{N}_{n_1,n_2}^{(33)}\tilde{N}_{p_1,p_2}^{(12)}
\prod_{j\in {\cal I}_1-\{1\}}\tilde{N}_{p_j,n_j}^{(13)}
\prod_{k\in{\cal I}_2-\{2\}}\tilde{N}_{p_k,n_k}^{(23)}.}}

Now let us compute the corresponding gauge theory diagrams
with an interaction vertex involving $\phi_{d_1},\phi_{d_2}$ or
$\bar{Z}$ in
${\cal O}^J_{(d_i,n_i)}$ and $\phi_{e_1},\phi_{e_2}$ or
$\bar{Z}$  in
${\cal T}^{J,y}_{(e_i,p_i);{\cal I}_1,{\cal I}_2}$. The result is:
\eqn\selfsemiDiff{ \eqalign{
\Gamma^{(1)}_{\{n_i\},\{p_i;{\cal I}_1,{\cal I}_2\}y}
\Big|_{1-2;1-2}^{\rm semi-nearest}
&={\sqrt{{y(1-y)\over J}}}\times\biggl[{1\over\sqrt{JJJ_1J_2}}
{g^2N\over 8\pi^2}(1-t_1^{J_1})(1-t_2^{J_1})\biggr]\cr
&\quad\times\prod_{j\in{\cal I}_1-\{1\}}{1\over\sqrt{JJ_1}}
\sum_{a=0}^{J_1-1}(t_j\bar{s}_j)^{a}
\prod_{k\in{\cal I}_2-\{2\}}{1\over\sqrt{JJ_2}}\sum_{b=0}^{J_2-1}
t_k^{J_1}(t_k\bar{s}_k)^{b}.}}
The interaction part of the diagram reduces to
\eqn\selffirstfactorDiff{\eqalign{
{1\over\sqrt{JJJ_1J_2}}{g^2N\over 8\pi^2}(1-t_1^{J_1})(1-t_2^{J_1})
&=-e^{\pi i(n_1+n_2)y}\lambda^\prime
{\sin(\pi n_1y)\sin(\pi n_2y)\over 2\pi^2\sqrt{y(1-y)}}\cr
&\simeq-2(-1)^{n_1+n_2+p_1}e^{\pi i(n_1+n_2)y}
\tilde{N}_{n_1,n_2}^{(33)}\tilde{N}_{p_1,p_2}^{(12)}
,}}
while the rest of the diagram, the free contraction contribution,
can be computed using \phasesum\ making
the phase disappear after imposing the level matching condition on
each trace.

Just as in the previous case, the Feynman diagrams we are considering
do not contribute to $\Gamma^{(0)}\Big|_l$ so that their contribution
to the matrix of anomalous dimensions in the string field theory basis
is given by
\eqn\selfGammaONEDiff{
\tilde{\Gamma}^{(1)}_{\{n_i\},\{p_i;{\cal I}_1,{\cal I}_2\}y}
\Bigr|_{1-2;1-2}^{\rm semi-nearest}\!\!\!\!
\simeq-2(-1)^n{\sqrt{{y(1-y)\over J}}}
\tilde{N}_{n_1,n_2}^{(33)}\tilde{N}_{p_1,p_2}^{(12)}
\prod_{j\in{\cal I}_1-\{1\}}\!\!\!\!\tilde{N}_{p_j,n_j}^{(13)}
\prod_{k\in{\cal I}_2-\{2\}}\!\!\!\!\tilde{N}_{p_k,n_k}^{(23)},}
which agrees with the string theory result \SFTamplifinb .

\newsec{Conclusion}

In this paper we have computed string interactions between string
states with an arbitrary number of scalar impurities. Using the
holographic map proposed in \GrossMH\GomisWI\ and the basis of gauge
theory states in \GomisWI\ we have exactly reproduced all string
amplitudes from gauge theory considerations. The calculations have
been carried up to ${\cal O}(g_2^2)$ for the case of two identical
impurities and to ${\cal O}(g_2)$ for arbitrary impurities. The
precise agreement found gives strong support to the validity of the
holographic map 
\identifiint\ and the basis of gauge theory states in \GomisWI.
The ${\cal O}(g_2^2)$ computation has been performed in
string field theory by truncating \RoibanXR\ by hand the allowed
intermediate
states. With this truncation we get precise agreement with the gauge
theory calculation. It is desirable to understand whether the
truncation is necessary.

While considering arbitrary string states, we have found that there is a
direct correspondence between the Feynman diagrams of gauge theory and
the string field theory Feynman diagrams that contribute to a given
amplitude. This diagrammatic correspondence is specially powerful when
we consider general string states, in which new classes of Feynman
diagrams appear as compared to the case with  two different
impurities. In particular we have shown which interaction vertex in
gauge theory corresponds to which string field theory vertex arising
from the action of the prefactor. Likewise the various Neumann
matrices in string theory have been derived from purely field
theoretic considerations as arising from various free contractions in
gauge theory.

The diagrammatic equivalence between gauge theory one loop diagrams
and string theory
    diagrams might be useful in deriving the duality, which is an 
important
open
problem. It would also be very desirable to represent the degrees of
freedom of the BMN sector
of ${\cal N}=4$ SYM by a complete theory, without any
truncation. 
 Holography strongly suggests that there should be a
quantum mechanical model which describes the BMN sector of ${\cal
N}=4$ SYM and at the same time captures all the physics of string
theory in the plane wave geometry. An important step in this direction
is the string bit model of Verlinde \VerlindeIG, but a suitable
non-abelian
generalization of it remains to be discovered.

\bigbreak\bigskip\bigskip\centerline{{\bf Acknowledgements}}\nobreak
We would like to thank Hirosi Ooguri and John Schwarz for useful
discussions.
J.~G.~is supported by a Sherman Fairchild Prize Fellowship. S.~M.~is
supported in part by JSPS Postdoctoral Fellowships for
Research Abroad H14-472.
This research was supported in part by the DOE grant
DE-FG03-92-ER40701.

\vfill

\appendix{A}{Asymptotic behavior of Neumann matrices}
In this appendix we present the asymptotic large $\mu$ behavior of
all Neumann matrices in the exponential basis.
These can be obtained from $(m,n\ne 0)$
\eqn\Ncossin{\tN^{(rs)}_{m,n}={1\over 2}
({\bar N}^{(rs)}_{|m|,|n|}-e(mn){\bar N}^{(rs)}_{-|m|,-|n|}),\qquad
\tN^{(rs)}_{m,0}={1\over\sqrt{2}}{\bar N}^{(rs)}_{|m|,0},\qquad
\tN^{(rs)}_{0,0}={\bar N}^{(rs)}_{0,0},}
where $e(m)=\hbox{sign}(m)$ and the asymptotic behavior of Neumann matrices in the $\cos$/$\sin$
basis in \HeZU:
\eqn\Nexp{\eqalign{
&{\tilde N}^{(11)}_{m,n}\simeq{(-1)^{m+n}\over 4\pi\mu y}\cr
&{\tilde N}^{(12)}_{m,n}\simeq{(-1)^{m+1}\over 4\pi\mu\sqrt{y(1-y)}}\cr
&{\tilde N}^{(22)}_{m,n}\simeq{1\over 4\pi\mu(1-y)}\cr
&{\tilde N}^{(13)}_{m,n}\simeq{(-1)^{m+n+1}\sin n\pi y
\over\pi\sqrt{y}(n-m/y)}\cr
&{\tilde N}^{(23)}_{m,n}\simeq{(-1)^n\sin n\pi y
\over\pi\sqrt{1-y}(n-m/(1-y))}\cr
&{\tilde N}^{(33)}_{m,n}\simeq
{(-1)^{m+n+1}\sin m\pi y\sin n\pi y\over\pi\mu}.}}

For the computation of the contact term, we also need ${\tilde F}^\pm$
in the exponential basis ($n\ne 0$)
\eqn\Fexp{{\tilde F}^\pm_{n(r)}={1\over\sqrt{2}}F^\pm_{|n|(r)},\qquad
{\tilde F}^\pm_{0(r)}=F^\pm_{0(r)}.}
and the scalar quantity $k$ and fermionic Neumann matrices
${\bar Y}$.
Using again the results in \HeZU, we have
\eqn\Ftilplus{\eqalign{
&{\tilde F}^+_{(1)n}\simeq(-1)^{n+1}\sqrt{\mu y}(1-y)\cr
&{\tilde F}^+_{(2)n}\simeq\sqrt{\mu(1-y)}y\cr
&{\tilde F}^+_{(3)n}\simeq{(-1)^{n+1}ny(1-y)\sin\pi ny\over\sqrt{\mu}},}}
\eqn\Ftilminus{\eqalign{
&{\tilde F}^-_{(1)n}\simeq{(-1)^{n+1}n(1-y)\over 2\sqrt{\mu y}}\cr
&{\tilde F}^-_{(2)n}\simeq{ny\over 2\sqrt{\mu(1-y)}}\cr
&{\tilde F}^-_{(3)n}
\simeq 2\sqrt{\mu}y(1-y)(-1)^{n+1}\sin\pi ny.}}
\eqn\R{
\!\!\!\!\!\!\!\!\!\!\!\!\!\!\!\!\!\!\!\!\!\!\!\!\!\!\!\!\!\!\!\!\!\!\!\!\!
\!\!\!\!\!\!\!\!\!\!\!\!\!\!\!\!\!\!\!\!\!\!\!\!\!\!\!\!\!\!\!\!\!\!\!\!\!
\!\!\!\!\!\!\!\!\!\!\!\!\!\!\!\!\!\!\!\!\!\!\!\!\!\!\!\!\!\!\!\!\!\!\!\!\!
\!\!\!
1-\mu y(1-y)k\simeq{1\over
4\pi\mu y(1-y)},}
\eqn\Ybar{{\bar Y}_0\simeq{1\over\sqrt{4\pi\mu y(1-y)}},\qquad
{\bar Y}_{n(1)}\simeq\sqrt{1-y\over 4\pi\mu}(-1)^{n+1},\qquad
{\bar Y}_{n(2)}\simeq\sqrt{y\over 4\pi\mu}.}

\appendix{B}{Matrix elements}
The definition of various matrices appearing in $G$ and $\Gamma$ on
the gauge theory calculation are given as follows.

\noindent{\bf $2$-impurity matrix elements}
($|m|\ne|n|, m\ne 0, n\ne 0,p\in Z, 0<y<1$)
\eqn\C{\eqalign{\bullet\quad&C_{n,py}=C_{py,n}
={y^{3/2}\sqrt{1-y}\over\sqrt{J}\pi^2}
{\sin^2(\pi n y)\over(p-ny)^2}\cr
&C_{n,y}=C_{y,n}=-{1\over\sqrt{J}\pi^2}{\sin^2(\pi ny)\over n^2}\cr
\bullet\quad&M^1_{n,n}={1\over 60}
-{1\over 24\pi^2n^2}+{7\over 16\pi^4n^4}\cr
&M^1_{n,-n}={1\over 48\pi^2n^2}+{35\over 128\pi^4n^4}\cr
&M^1_{n,m}={1\over 12\pi^2(n-m)^2}-{1\over 8\pi^4(n-m)^4}
+{1\over 4\pi^4n^2m^2}+{1\over 8\pi^4nm(n-m)^2}\cr
\bullet\quad&\Gamma^{(1)}_{n,py}=\Gamma^{(1)}_{py,n}
=\lambda'\left({p^2\over y^2}-{pn\over y}+n^2\right)C_{n,py}\cr
&\Gamma^{(1)}_{n,y}=\Gamma^{(1)}_{y,n}=\lambda'n^2C_{n,y}\cr
&\Gamma^{(2)}_{n,m}=\lambda'
nmM^1_{n,m}+{1\over 8\pi^2}{\cal D}^1_{n,m}\cr
\bullet\quad&{\cal D}^1_{n,n}={\cal D}^1_{n,-n}
=\lambda'\left({2\over 3}+{5\over\pi^2n^2}\right)\cr
&{\cal D}^1_{n,m}=\lambda'\left({2\over
3}+{2\over\pi^2n^2}+{2\over\pi^2m^2}\right)\cr
\bullet\quad&B_{n,n}={1\over 3}+{5\over 2\pi^2n^2}\cr
&B_{n,-n}=-{15\over 8\pi^2n^2}\cr
&B_{n,m}={3\over 2\pi^2mn}+{1\over 2\pi^2(m-n)^2}}}

\noindent{\bf $n$-impurity matrix elements}

\eqn\nMatrix{\eqalign{
\bullet\quad &G^{(1)}_{\{n_i\},\{p_i;{\cal I}_1,{\cal I}_2\}y}
=G^{(1)}_{\{p_i;{\cal I}_1,{\cal I}_2\}y,\{n_i\}}\cr
&= (-1)^{n+\sum_{k\in {\cal I}_2}n_k}
{ \sqrt{y^{n_1+1}}\sqrt{(1-y)^{n_2+1}}
    \over\sqrt{J}}  \prod_{j\in {\cal I}_1} {\sin(\pi n_jy)\over \pi 
(p_j-n_jy)}
   \prod_{k\in {\cal I}_2} {\sin(\pi n_k(1-y))\over \pi (p_k-n_k(1-y))} , 
\cr
\bullet\quad &\Gamma^{(1)}_{\{n_i\},\{p_i;{\cal I}_1,{\cal I}_2\}y}=
\Gamma^{(1)}_{\{p_i;{\cal I}_1,{\cal I}_2\}y,\{n_i\}} \cr
&= {\lambda^\prime \over2}
\left[ \sum_{j\in{\cal I}_1}\left( \left(n_j-{p_j\over 
y}\right)^2+n_j{p_j\over
y}\right)+\sum_{k\in {\cal I}_2}\left( \left(n_k-{p_k\over 
1-y}\right)^2  +n_k{p_k\over
1-y}\right) \right]\cr
&\;\;\;\; \times  G^{(1)}_{\{n_i\},\{p_i;{\cal I}_1,{\cal I}_2\}y}.
}}
where $n_1=|{\cal I}_1|,\;\;n_2=|{\cal I}_2|$.

\appendix{C}{Calculation of supersymmetry charge matrix elements}
In this appendix we shall show explicitly how to reduce the
supersymmetry vertex in \PankiewiczTG\ to our simple formula
\SUSYelement\ when we assume the external state to have two bosonic
impurities and the intermediate state to have one bosonic and one
fermionic impurity. See also \RoibanXR.

The Hamiltonian and supersymmetry charge
vertices in \PankiewiczTG\ are given by:
\eqn\hamilton{\eqalign{|H_3\ra
&=c\Bigl((1+\mu\al k)(K_+^i-K_-^i)(K_+^j+K_-^j)
-\mu\al\delta^{ij}\Bigr)v_{ij}(Y)E_aE_bE_{b0}|0\ra,\cr
|Q_{3\da}\ra&=c(1+\mu\al k)^{1/2}(K_+^i-K_-^i)
s_\da^i(Y)E_aE_bE_{b0}|0\ra,\cr
|{\bar Q}_{3\da}\ra&=c(1+\mu\al k)^{1/2}(K_+^i+K_-^i)
{\tilde s}_\da^i(Y)E_aE_bE_{b0}|0\ra.}}
Various constituents of the prefactor, $K_\pm^i$, $v^{ij}$
$s^i_\da=-i\sqrt{2}(\eta s^i_{1\da}+{\bar\eta}s^i_{2\da})$ and
${\tilde s}^i_\da=i\sqrt{2}({\bar\eta}s^i_{1\da}+\eta s^i_{2\da})$
are given as
\eqn\K{\eqalign{K_+^i&=\sum_{r=1}^3\sum_{m=-\infty}^\infty
{\tilde F}^+_{m(r)}\al_{m(r)}^{i\dagger},\cr
K_-^i&=\sum_{r=1}^3\sum_{m=-\infty}^\infty
{\tilde F}^-_{m(r)}\al_{m(r)}^{i\dagger},}}
\eqn\v{\eqalign{v^{ij}&=\delta^{ij}
+{1\over 4!\al^2}t^{ij}_{abcd}Y^aY^bY^cY^d
+{1\over 8!\al^4}\delta^{ij}\epsilon_{abcdefgh}Y^a\cdots Y^h\cr
&\qquad+{1\over 2!\al}\gamma^{ij}_{ab}Y^aY^b
+{1\over 2!6!\al^3}\gamma^{ij}_{ab}
\epsilon^{ab}{}_{cdefgh}Y^c\cdots Y^h,\cr
{1\over\sqrt{2}}s^i_{1\da}&=\gamma^i_{a\da}Y^a
+{1\over 3!5!\al^2}u^i_{abc\da}\epsilon^{abc}{}_{defgh}
Y^d\cdots Y^h,\cr
{1\over\sqrt{2}}s^i_{2\da}&=-{1\over 3!\al}u^i_{abc\da}Y^aY^bY^c
+{1\over 7!\al^3}\gamma^i_{a\da}\epsilon^{a}{}_{bcdefgh}
Y^b\cdots Y^h,}}
where $Y^a$ reads
\eqn\Y{Y^a=\sqrt{2}Y_0(\al_{(1)}\lambda_{(2)}^a
-\al_{(2)}\lambda_{(1)}^a)
+\sum_{r=1}^3\sum_{m=1}^\infty Y_{m(r)}b_{m(r)}^{a\dagger},}
with
\eqn\lamb{\lambda_{(r)}^a=\sqrt{\al_{(r)}\over 2}
\pmatrix{b_{0(r)}^{a\dagger}\cr b_{0(r)}^a}\quad (r=1,2),\qquad\qquad
\lambda_{(3)}^a={1\over\sqrt{2}}
\pmatrix{b_{0(3)}^a\cr b_{0(3)}^{a\dagger}},}
and
\eqn\Yzero{Y_0={\bar Y}_0\pmatrix{1&\cr&0}
+{1\over{\bar Y}_0}\pmatrix{0&\cr&1},}
\eqn\Yposi{Y_{n(1)}={\bar Y}_{n(1)}\pmatrix{1&\cr&0},\qquad
Y_{n(2)}={\bar Y}_{n(2)}\pmatrix{1&\cr&0}.}
Note that in the matrix representation of \lamb , \Yzero\ and \Yposi ,
the upper(left) entries denote the components with spinor indices
$a=1,\cdots,4$, while the lower(right) ones denote $a=5,\cdots,8$.
Also $E_a$, $E_b$ and $E_{b0}$ come from the overlapping condition
of bosonic modes, fermionic non-zero modes and fermionic zero modes,
respectively
\eqn\Ea{E_a=\exp\biggl({1\over 2}
\sum_{r,s=1}^3\sum_{m,n=-\infty}^\infty
\al_{(r)m}^{i\dagger}{\tilde N}^{(rs)}_{mn}\al_{(s)n}^{i\dagger}
\biggr),}
\eqn\Ebzero{E_{b0}={1\over 2^4}\prod_{a=1}^4
(\sqrt{\al_{(1)}}b_{(1)}^{a\dagger}+\sqrt{\al_{(2)}}b_{(2)}^{a\dagger}
+b_{(3)}^a)
\prod_{b=5}^8(\sqrt{\al_{(1)}}b_{(1)}^b+\sqrt{\al_{(2)}}b_{(2)}^b
+b_{(3)}^{b\dagger}),}
and the explicit expression of $E_b$ is not necessary in our
analysis.
Finally the ``ground'' state $|0\ra$ is related to the
``vacuum'' state with the lowest energy by:
\eqn\zero{|0\ra=\prod_{a=5}^8b_{(1)}^{a\dagger}
\prod_{b=5}^8b_{(2)}^{b\dagger}
\prod_{c=1}^4b_{(3)}^{c\dagger}|\vac\ra.}

Now we would like to calculate the supersymmetry charge matrix
elements
\eqn\Qelem{Q_{n,m(s)}=\la\vac|\al_{n(3)}^i\al_{-n(3)}^i
\al_{m(s)}^k{1\over\sqrt{2}}(b_{m(s)}^d-ie(m)b_{-m(s)}^d)|Q_\da\ra,}
where we assume the external states to be two bosonic impurity states
and the intermediate states to be states with one bosonic and one
fermionic impurity.
The supersymmetry charge matrix elements under this assumption
will be greatly simplified.
We find the $(Y)^1$ and $(Y)^7$ terms in \v\ vanish and the
$(Y)^3$ and $(Y)^5$ terms reduce to \SUSYelement.

The typical matrix element of the supersymmetry charge \Qelem\
is
\eqn\typical{\la\vac|\al_{n(3)}^i\al_{-n(3)}^i\al_{m(s)}^k
{1\over\sqrt{2}}(b_{m(s)}^d-ie(m)b_{-m(s)}^d)K^\pm Y^\ell
E_aE_bE_{b0}|0\ra,}
where $\ell$ denotes the number of fermions in the expression \v.
First of all, let us concentrate on the zero mode $b_{0(3)}$
operators.
Since we only have $b_{0(3)}$ in $E_{b0}$ and $|0\ra$, all the
$b_{0(3)}$ operators should cancel out to obtain a non-vanishing
contribution.
The only possibility is that $b_{0(3)}^a$ ($a=1,\cdots,4$) in $E_{b0}$
cancels those in $|0\ra$ and we never use $b_{0(3)}^b$
($b=5,\cdots,8$) in $E_{b0}$.
Using this fact, our typical matrix elements become
\eqn\typic{\la\vac|\al_{n(3)}^i\al_{-n(3)}^i\al_{m(s)}^k
{1\over\sqrt{2}}(b_{m(s)}^d-ie(m)b_{-m(s)}^d)K^\pm Y^\ell
E_aE_{b0}\prod_{a=5}^8b_{0(1,2)}^{a\dagger}|\vac\ra,}
with $b_{0(1,2)}^a$ meaning $b_{0(1)}^a$ or $b_{0(2)}^a$.
Next, let us concentrate on the zero mode $b_{0(1,2)}$ operators.
In case of $\ell=1$, we will not have enough annihilation operators to
cancel all the leftover zero modes in $|0\ra$.
In case  $\ell=7$, we use four of $b(Y)^7$ to cancel the zero modes.
But the rest must all be the creation operators and now we have too
many of them.
If $\ell=3$, exactly four operators in $b(Y)^3$ are used to cancel the
leftover in $|0\ra$.
If $\ell=5$, four in $b(Y)^5$ are used to cancel.
Since $Y$ does not have both the creation operators and annihilation
ones for the same operator, two of the $Y$'s cannot cancel each other.
Therefore we have to choose four operators in $Y$ to cancel the
creation operators in $|0\ra$ and let the remaining $Y$'s be cancelled by
the $b$ of the intermediate state.

For the $(Y)^3$ term, only the zero modes contribute:
\eqn\susyzero{\sim{\bar Y}_0
({\tilde F}^\pm_{(1)0}{\tilde N}^{(33)}_{n,-n}
+{\tilde F}^\pm_{(3)n}{\tilde N}^{(13)}_{0,-n}
+{\tilde F}^\pm_{(3)-n}{\tilde N}^{(13)}_{0,n}).}
For the $(Y)^5$ term, besides the zero modes contribution, the non-zero
modes also contribute as:
\eqn\susynonzero{\sim{{\bar Y}_{m(1)}\over\sqrt{2}}
({\tilde F}^\pm_{(1)m}{\tilde N}^{(33)}_{n,-n}
+{\tilde F}^\pm_{(3)n}{\tilde N}^{(13)}_{m,-n}
+{\tilde F}^\pm_{(3)-n}{\tilde N}^{(13)}_{m,n}).}
Using the large $\mu$ behavior of various Neumann coefficients in
Appendix A, we find that ${\tilde F}^-_{(3)}{\tilde N}^{(13)}$ gives
the leading contribution.
Besides, from the symmetry of the Neumann coefficients, we have
\eqn\vanish{{\tilde F}^-_{(3)n}{\tilde N}^{(13)}_{0,-n}
+{\tilde F}^-_{(3)-n}{\tilde N}^{(13)}_{0,n}\simeq 0.}
Therefore the only relevant matrix element of the supersymmetry
charge comes from $m\ne 0$.

For the analysis of the normalization of the contact term, let us be
careful about the overall factor.
Since
\eqn\onedirect{(\al_{(1)}\sqrt{\al_{(2)}}b_{0(2)}^a
-\al_{(2)}\sqrt{\al_{(1)}}b_{0(1)}^a)
(\sqrt{\al_{(1)}}b_{0(1)}^a+\sqrt{\al_{(2)}}b_{0(2)}^a)
b_{0(1)}^{a\dagger}b_{0(2)}^{a\dagger}|\vac\ra
=-\sqrt{\al_{(1)}\al_{(2)}}|\vac\ra,}
for $a=5,\cdots,8$, from the cancellation of the fermionic zero modes
we have an extra factor of
\eqn\factor{c_0=\biggl({\sqrt{\al_{(1)}\al_{(2)}}
\over 2{\bar Y}_0}\biggr)^4.}
Taking it into account, we find that the only non-trivial contribution
is:
\eqn\barQsamenonzero{Q_{n,m(s)}
=i\eta{2cc_0\over\al^2}\sqrt{1+\mu\al k}
u^i_{abc\da}\delta^{abcd}_{1234}{{\bar Y}_{m(1)}\over\sqrt{2}}
({\tilde F}^-_{(3)-n}{\tilde N}^{(s3)}_{m,n}
+{\tilde F}^-_{(3)n}{\tilde N}^{(s3)}_{m,-n}).}

To fix the overall normalization, let us compare the supersymmetry
charge matrix element with the Hamiltonian.
If we restrict the external states to be purely bosonic ones, we also
have the same fermionic zero mode factor $c_0$ in the Hamiltonian
matrix element:
\eqn\ham{\eqalign{|H_3\ra&={cc_0\over\al^2}(1+\mu\al k)
(K^i_+-K^i_-)(K^j_++K^j_-)t^{ij}_{5678}E_a|\vac\ra\cr
&={2cc_0\over\al^2}\biggl(-{y(1-y)\over 2}\biggr)
\sum_{r=1}^3\sum_{n=-\infty}^\infty
{\omega_{n(r)}\over\al_{(r)}}\al_{n(r)}^{i\dagger}\al_{-n(r)}^i
E_a|\vac\ra.}}
In the final step, we have used the formula derived in \LeeVZ.
Comparing our final expression with \Hone\ whose normalization factor
was determined  in \PearsonZS\GomisWI\ by comparing the string field
theory  result
with a gauge theory computation,
   we
find that
\eqn\norm{{2cc_0\over\al^2}=1.}

\appendix{D}{Formulas for calculating the contact term}
The necessary summation and integration we need to calculate the
contact term are the following ones:
\eqn\NN{\eqalign{\sum_{l=-\infty}^\infty
\tN^{(13)}_{l,n}\tN^{(13)}_{l,m}
&={(-1)^{m+n}\sin(n-m)\pi y\over\pi(n-m)},\cr
\sum_{l=-\infty}^\infty
\tN^{(23)}_{l,n}\tN^{(23)}_{l,m}
&={\sin(n-m)\pi(1-y)\over\pi(n-m)}.}}
Also,
\eqn\integrate{\eqalign{&\int_0^1dy(-1)^{m+n}
{\sin\pi my\sin\pi ny\over\pi^2}
\biggl\{(-1)^{l+n}\biggl({\sin\pi(m-n)y\over\pi(m-n)}
-{\sin\pi(m+n)y\over\pi(m+n)}\biggr)(1-y)\cr
&\qquad\qquad\qquad\qquad\qquad\qquad
+\biggl({\sin\pi(m-n)(1-y)\over\pi(m-n)}
-{\sin\pi(m+n)(1-y)\over\pi(m+n)}\biggr)y\biggr\}\cr
&\qquad\qquad\qquad
={1\over 4\pi^4(m-n)^2}+{1\over 4\pi^4(m+n)^2},\cr
&\int_0^1dy{\sin^2\pi ny\over\pi^2}
\biggl\{\biggl(y-{\sin 2\pi ny\over 2\pi n}\biggr)(1-y)
+\biggl((1-y)-{\sin 2\pi n(1-y)\over 2\pi n}\biggr)y\biggr\}\cr
&\qquad\qquad\qquad={1\over 2\pi^2}
\biggl({1\over 3}+{5\over 8\pi^2n^2}\biggr).}}

\appendix{E}{$\tilde\Gamma^{(2)}$ computation}
In this appendix, we explain the details of the computation of
${\tilde\Gamma}^{(2)}$ matrix elements for the operators with two
impurities in the same direction, as discussed in section 3.
The following identity will be useful throughout the  computation:
\eqn\sym{C_{n,py}=C_{-n,-py}.}
As in \anombasis, ${\tilde\Gamma}^{(2)}$ is given by:
\eqn\tildeGamma{{\tilde\Gamma}^{(2)}=\Gamma^{(2)}
-{1\over 2}\{G^{(2)},\Gamma^{(0)}\}
-{1\over 2}\{G^{(1)},\Gamma^{(1)}\}
+{3\over 8}\{G^{(1)2},\Gamma^{(0)}\}
+{1\over 4}G^{(1)}\Gamma^{(0)}G^{(1)}.}
Here we shall compute all the terms and show that $\tilde\Gamma^{(2)}$
reduces to \tildeGammatwoiijj.

Our strategy is to split each matrix element in \anomalous\ into a
part proportional to $\delta_{ij}$ and a part coming  from extra
diagrams.
More precisely, we have
\eqn\splitting{\eqalign{
\Gamma^{(1)}_{iin,jjqz}
&=\delta_{ij}\bigl(\Gamma^{(1)}_{n,qz}+\Gamma^{(1)}_{-n,qz}\bigr)
+\delta\Gamma^{(1)}_{n,qz},\cr
\Gamma^{(1)}_{iin,jjz}
&=\delta_{ij}\bigl(\Gamma^{(1)}_{n,z}+\Gamma^{(1)}_{-n,z}\bigr)
+\delta\Gamma^{(1)}_{n,z},\cr
\Gamma^{(2)}_{iin,jjm}
&=\delta_{ij}
\bigl(\Gamma^{(2)}_{n,m}+\Gamma^{(2)}_{n,-m}\bigr)
+\delta\Gamma^{(2)}_{n,m},}}
with
\eqn\deltaGamma{\eqalign{
\delta\Gamma^{(1)}_{n,qz}&=-{1\over2}\Gamma^{(1)}_{n,0z},\cr
\delta\Gamma^{(1)}_{n,z}&=-{1\over2}\Gamma^{(1)}_{n,z},\cr
\delta\Gamma^{(2)}_{n,m}&=-{1\over 16\pi^2}{\cal D}^1_{n,m}.}}

As a preliminary computation let us consider $(G^{(1)})^2$:
\eqn\Gonesq{\eqalign{(G^{(1)})^2&=J\int_0^1dy
\Biggl(\sum_{p=1}^\infty\bigl(C_{n,py}+C_{n,-py}\bigr)
\bigl(C_{py,m}+C_{-py,m}\bigr)
+2C_{n,0y}C_{0y,m}\Biggr)\cr
&\qquad\qquad+J\int_0^{1/2}dy2C_{n,y}2C_{y,m}\cr
&=J\int_0^1dy\sum_{p=-\infty}^\infty
\Bigl(C_{n,py}C_{py,m}+C_{n,py}C_{py,-m}\Bigr)
+2J\int_0^1dyC_{n,y}C_{y,m}\cr
&={1\over 2}\bigl(M^1_{n,m}+M^1_{n,-m}\bigr).}}
Here we have to be careful about the extra normalization factor
$1/\sqrt{2}$ for zero modes as explained around \anomalous.
Note that originally in the first line we sum  only over positive
integers the product of two terms.
One of the terms is the product of two contributions with opposite
worldsheet momentum.
But with the help of \sym, we can rewrite these cross terms into the
summation of two terms over all the integers, with still one of them
carrying the reversed {\it external} worldsheet momentum as in the
second equation in \Gonesq.
Since one of two terms is identical to the one arising for operators
with two
impurities in different directions, we can perform the summation and
integration easily and add the other term by reversing the external
worldsheet momentum.
This kind of mechanism happens everywhere, also in the computation of
$\tilde\Gamma^{(2)}$.
Therefore, the naive expectation of $\tilde\Gamma^{(2)}$ is obtained by
adding a term with the external worldsheet momentum reversed:
\eqn\ignore{\tilde\Gamma^{(2)}_{iin,jjm}=\delta_{ij}
\bigl(\tilde\Gamma^{(2)}_{n,m}+\tilde\Gamma^{(2)}_{n,-m}\bigr)
=\delta_{ij}{1\over 16\pi^2}\bigl(B_{n,m}+B_{n,-m}\bigr).}
The only point we have to be careful with is whether \deltaGamma\ will
give a non-trivial contribution.

Let us postpone the effect of \deltaGamma\ and concentrate on the
dominant contribution to see whether the results have an additional
contribution of reversing the worldsheet momentum, as compared to the
case of operators with two impurities in different directions.
Now it is quite trivial to calculate terms involving $\Gamma^{(0)}$ in
\tildeGamma\
such as $\{(G^{(1)})^2,\Gamma^{(0)}\}$, $G^{(1)}\Gamma^{(0)}G^{(1)}$
and $\{G^{(2)},\Gamma^{(0)}\}$. They are given by:
\eqn\Gammazeroth{\eqalign{
\{(G^{(1)})^2,\Gamma^{(0)}\}
&={n^2+m^2\over 2}(M^1_{n,m}+M^1_{n,-m}),\cr
G^{(1)}\Gamma^{(0)}G^{(1)}
&=J\int_0^1dy\sum_{p=0}^\infty(C_{n,py}+C_{n,-py})
{p^2\over y^2}(C_{py,m}+C_{-py,m})\cr
&=J\int_0^1dy\sum_{p=-\infty}^\infty
(C_{n,py}{p^2\over y^2}C_{py,m}+C_{n,py}{p^2\over y^2}C_{py,-m}),\cr
\{G^{(2)},\Gamma^{(0)}\}
&=(n^2+m^2)(M^1_{n,m}+M^1_{n,-m}).}}
Let us turn to the term involving $\Gamma^{(1)}$ in \tildeGamma, but
ignoring the effect of
\deltaGamma. It is given by:
\eqn\GoneGamone{\eqalign{
&\{G^{(1)},(\Gamma^{(1)}-\delta\Gamma^{(1)})\}\cr
&=J\int_0^1dy\sum_{p=1}^\infty
(C_{n,py}+C_{n,-py})(\Gamma^1_{py,m}+\Gamma^1_{-py,m})
+{1\over2}(C_{n,0y}+C_{n,0y})(\Gamma^1_{0y,m}+\Gamma^1_{0y,m})\cr
&\quad+J\int_0^1dy\sum_{p=1}^\infty
(\Gamma^1_{n,py}+\Gamma^1_{n,-py})(C_{py,m}+C_{-py,m})
+{1\over2}(\Gamma^1_{n,0y}+\Gamma^1_{n,0y})(C_{0y,m}+C_{0y,m})\cr
&\quad+J\int_0^{1/2}dy
(4C_{n,y}\Gamma^1_{y,m}+4\Gamma^1_{n,y}C_{y,m})\cr
&=J\int_0^1dy\Biggl\{\sum_{p=-\infty}^\infty
(C_{n,py}\Gamma^1_{py,m}+\Gamma^1_{n,py}C_{py,m})
+(C_{n,y}\Gamma^1_{y,m}+\Gamma^1_{n,y}C_{y,m})\Biggr\}\cr
&\quad+J\int_0^1dy\Biggl\{\sum_{p=-\infty}^\infty
(C_{n,py}\Gamma^1_{py,-m}+\Gamma^1_{n,py}C_{py,-m})
+(C_{n,y}\Gamma^1_{y,-m}+\Gamma^1_{n,y}C_{y,-m})\Biggr\}.}}
Also if we ignore the effect of \deltaGamma, $\Gamma^{(2)}$ also has
the same additional contribution, as seen in \splitting.
As promised, all the results come  paired with  $(n,m)$ and $(n,-m)$,
where the first group of terms adds up to give the same result as for
the case of two different impurities.

Now let us consider the contribution of $\delta\Gamma$'s to
$\tilde\Gamma^{(2)}$
\eqn\deltatildeGamma{
\delta\tilde\Gamma^{(2)}
=\delta\Gamma^{(2)}-{1\over 2}\{G^{(1)},\delta\Gamma^{(1)}\},}
that we have not taken into account so far.
We can compute the second term as before:
\eqn\secondterm{\eqalign{
&\{G^{(1)},\delta\Gamma^{(1)}\}_{iin,jjm}\cr
&=J\int_0^1dy\left\{\sum_{p=1}^\infty
(C_{n,py}+C_{n,-py})\biggl(-{1\over2}\Gamma^{(1)}_{0y,m}\biggr)
+C_{n,0y}\biggl(-{1\over2}\Gamma^{(1)}_{0y,m}\biggr)\right\}\cr
&\quad+J\int_0^1dy\left\{\sum_{p=1}^\infty
\biggl(-{1\over2}\Gamma^{(1)}_{n,0y}\biggr)(C_{py,m}+C_{-py,m})
+\biggl(-{1\over2}\Gamma^{(1)}_{n,0y}\biggr)C_{0y,m}\right\}\cr
&\quad+J\int_0^{1\over2}dy
(2C_{n,y})\biggl(-{1\over2}\Gamma^{(1)}_{y,m}\biggr)
+\biggl(-{1\over2}\Gamma^{(1)}_{n,y}\biggr)(2C_{y,n})\cr
&=-{J\over2}\int_0^1dy\sum_{p=-\infty}^\infty
\Bigl(C_{n,py}\Gamma^1_{0y,m}+\Gamma^1_{n,0y}C_{py,m}\Bigr)
-{n^2+m^2\over 2}J\int_0^1dyC_{n,y}C_{y,m}.}}
Using the formula,
\eqn\formulae{
J\int_0^1dy\sum_{p=-\infty}^\infty C_{n,py}\Gamma^1_{0y,m}
={1\over 12\pi^2}\biggl(1+{3\over\pi^2m^2}\biggr),  }
and the summation formula in the appendix of \GomisWI, we obtain
\eqn\secondtermresult{\{G^{(1)},\delta\Gamma^{(1)}\}_{iin,jjm}
=-{1\over 8\pi^2}{\cal D}^1_{n,m},}
which precisely cancels $\delta\Gamma^{(2)}$:
\eqn\answer{\delta\tilde\Gamma^{(2)}=0.}
Therefore we find that \ignore\ is exact.
In section 3 and appendices B and C, we saw that this result is
correctly reproduced from the contact term calculation in string field
theory.

\appendix{F}{Anomalous dimension of the singlet operators}
In this appendix we shall calculate the anomalous dimension of the
operator with two impurities in the same direction, using the
perturbation theory.
This calculation has essentially been done in \BeisertBB\ by
diagonalizing the matrix of two-point functions in the BMN basis.
Here, we perform the calculation using the string field theory basis
and it serves as  a
   consistency check of the
evaluation of $\tilde\Gamma^{(2)}$ in the previous appendix.

In perturbation theory the eigenvalue at $O(g_2^2)$ is given by:
\eqn\perturb{\Delta^{(2)}
=J\int_0^1dy \sum_{p=0}^\infty\sum_{j=1}^4
{({\tilde\Gamma}_{ii;n,jj;py}^{(1)})^2\over n^2-p^2/y^2}
+J\int_0^{1/2}dy\sum_{j=1}^4
{({\tilde\Gamma}_{ii;n,jj;y}^{(1)})^2\over n^2}
+{\tilde\Gamma}_{ii;n,ii;n}^{(2)}.}
Using the following relations
\eqn\relations{\eqalign{{\tilde\Gamma}^{(1)}_{n,py}&={1\over 2}
\Gamma^{(1)}_{n,0y},\cr
{\tilde\Gamma}^{(1)}_{n,y}&={1\over 2}\Gamma^{(1)}_{n,y},}}
\perturb\ can be rewritten as:
\eqn\simplifyperturb{\Delta^{(2)}
={J\over 2}\int_0^1dy \sum_{p=-\infty}^\infty
{({\Gamma}_{n,0y}^{(1)})^2\over n^2-p^2/y^2}
+{J\over2}\int_0^{1}dy{({\Gamma}_{n,y}^{(1)})^2\over n^2}
+{\tilde\Gamma}_{ii;n,ii;n}^{(2)}.}
Now let us proceed to evaluate each term. The first term is
\eqn\firstterm{{1\over 2\pi^4}\int_0^1dy{1-y\over y}
\sin^4(\pi ny)\sum_{p=-\infty}^\infty{1\over n^2-p^2/y^2}
={3\over 64\pi^4n^2},}
and the second term is simply reduced to the integration of
$(C_{n,y})^2$,
whose result can be found in \GomisWI:
\eqn\secondterm{
{n^2\over 2}J\int_0^1dy C_{n,y}^2={3\over 16\pi^4n^2}.}

Consequently, the anomalous dimension of the singlet operator is
\eqn\anomalousdim{\Delta^{(2)}
={3\over 64\pi^4n^2}+{3\over 16\pi^4n^2}
+{1\over 16\pi^2}\bigl(B_{n,n}+B_{n,-n}\bigr)
={1\over 16\pi^2}\biggl({1\over 3}+{35\over 8\pi^2n^2}\biggr),}
which as explained in \BeisertBB\ConstableVQ\ is
   the same as the operators transforming in the ${\bf 6}$ and ${\bf 9}$
representations of $SO(4)$.

\appendix{G}{BMN operators in complex field notation}
In the main text, we have used real scalar field notation to define BMN
operators with arbitrary combination of impurities. In this case,
we have
four kinds of scalar impurities $\phi_1,\phi_2,\phi_3,\phi_4$ which
can be
inserted, and a subtlety arises when two identical impurities collide.
In this appendix, we study the same problem in the complex scalar
field formulation. In this formulation, there are also four kinds of
impurities $\Phi,\bar{\Phi},\Psi,\bar{\Psi}$. First, we want to see if
BMN operators with anti-holomorphic insertions are well defined in the
BMN limit. For example, let us consider $\Phi$ and $\bar{\Psi}$
insertions:
\eqn\PhibarPsi{{\cal O}^J_{\Phi\bar{\Psi},n}
={1\over\sqrt{JN^{J+2}}}\sum_{l=0}^Je^{2\pi iln/J}
\Tr\left(\Phi Z^l \bar{\Psi} Z^{J-l}\right).}
{}From the original Lagrangian of ${\cal N}=4$ SYM theory, it is easy
to see that there is a symmetry which maps $\phi_4$ to $-\phi_4$,
thereby transforming $\Psi$ to $\bar{\Psi}$ without changing $Z$ and
$\Phi$. From the ten-dimensional ${\cal N}=1$ SYM viewpoint, it is
just the reflection along one of the internal directions. In terms of
an
${\cal N}=1$ superfield formulation of ${\cal N}=4$ SYM, it is
equivalent to treating $Z,\Phi,\bar{\Psi}$ as chiral superfields
instead of $Z,\Phi,\Psi$. (The original D-term potential and F-term
potential should regroup to give the same form of D-term and F-term
potential in terms of $Z,\Phi,\bar{\Psi}$.) Therefore, the Feynman
diagram computation is identical to that for BMN operators with $\Phi$
and $\Psi$ insertions, as it should be because in terms of the real
scalar representation the four impurities are equivalent as far as
same impurities do not collide. Hence, we conclude that the four
complex impurities are equivalent in the dilute gas approximation.

Now let us think about the subtlety arising when two impurities
collide. In the real scalar representation, only when two same
impurities collide we had to add an extra term with $\bar{Z}$
insertion. In the complex scalar representation, this extra term is
necessary only when $\Phi$ and $\bar{\Phi}$ collide or $\Psi$ and
$\bar{\Psi}$ collide. This can be understood from the action of
$R$-symmetry generators on the BMN operators.(See also \ParnachevKK.)
Let us denote the $R$-symmetry generator of the rotation on
$\phi_i$-$\phi_j$ plane by $R_{ij}=-R_{ji}$. More precisely,
\eqn\Rijaction{R_{ij}\cdot\phi_j=\phi_i,
\qquad R_{ij}\cdot\phi_i=-\phi_j.}
Then define
\eqn\RPhiZ{R_{\Phi Z}
={1\over 2}\left(R_{15}+R_{26}+iR_{25}-iR_{16}\right),\qquad
R_{\bar{\Phi}Z}={1\over 2}\left(R_{15}-R_{26}-iR_{25}-iR_{16}\right).}
Their actions are given as
\eqn\RPhiZaction{\eqalign{&R_{\Phi Z}\cdot Z=\Phi,\qquad
R_{\Phi Z}\cdot\bar{\Phi}=-\bar{Z},\qquad
R_{\Phi Z}\cdot\bar{Z}=R_{\Phi Z}\cdot\Phi=0,\cr
&R_{\bar{\Phi}Z}\cdot Z=\bar{\Phi},\qquad
R_{\bar{\Phi}Z}\cdot\Phi=-\bar{Z},\qquad
R_{\bar{\Phi}Z}\cdot\bar{Z}=R_{\bar{\Phi}Z}\cdot\bar{\Phi}=0,}}
and likewise for $R_{\Psi Z}$ and $R_{\bar{\Psi}Z}$. BPS BMN operators
can be obtained by acting these generators successively on the vacuum
operator $\Tr(Z^J)$. For example, if we want to insert $\Phi$ and
$\Psi$, we act with $R_{\Phi Z}$ and $R_{\Psi Z}$ on $\Tr(Z^{J+2})$,
\eqn\PhiPsi{  R_{\Psi Z} \cdot\left(R_{\Phi Z}\cdot
\Tr(Z^{J+2})\right)=R_{\Psi Z} \cdot\left(\sum_{l=0}^{J+1} \Tr(Z^l \Phi
Z^{J+1-l})\right)=(J+2)\sum_{l=0}^J \Tr(\Phi Z^l \Psi Z^{J-l}).}
Since $R_{\Psi Z} \cdot \Phi =0$, we don't have any extra term arising
when $R_{\Psi Z}$ acts on $\Phi$. It is also the case when we insert two
$\Phi$'s because $R_{\Phi Z} \cdot \Phi =0$. Now let us consider $\Phi$
and $\bar{\Phi}$ insertions.
\eqn\PhibarPhi{\eqalign{
&R_{\bar{\Phi}Z}\cdot\left(R_{\Phi Z}\cdot\Tr(Z^{J+2})\right)
=R_{\bar{\Phi}Z}\cdot\left(\sum_{l=0}^{J+1}
\Tr(Z^l\Phi Z^{J+1-l})\right)\cr
&\qquad=(J+2)\left(\sum_{l=0}^J\Tr(\Phi Z^l\bar{\Phi}Z^{J-l})
-\Tr(\bar{Z}Z^{J+1})\right).}}
The $\bar{Z}$ term arises when $R_{\bar{\Phi} Z}$ acts on $\Phi$, in
other words, when $\Phi$ and $\bar{\Phi}$ ``collide''. We conclude that
only when holomorphic and antiholomorphic insertions of the same kind
collide, we need to add an extra $\bar{Z}$ term. From this
consideration, we can also learn that no extra term is necessary when
$\bar{Z}$ collides with the four impurities because all the four
generators annihilate $\bar{Z}$. For example, when $R_{\Phi Z}$ acts on
$\Tr(\bar{Z}Z^{J+1})$,
\eqn\RPhiZ{ R_{\Phi Z} \cdot \Tr(\bar{Z}Z^{J+1}) = \sum_{l=0}^J
\Tr(\bar{Z} Z^l \Phi Z^{J-l}).}
This implies that we don't have to worry about collision of more than
two impurities. In general, we have only to take care of holomorphic and
antiholomorphic impurities of the same kind pairwise.

\appendix{H}{Off-shell representation of BMN operators}
In this appendix, we carefully define ``on-shell'' and ``off-shell''
representations of BMN operators which are introduced in the main
text. (See also \ConstableHW.) Here by shell we mean the level matching 
condition shell, which states that the sum of all worldsheet momentum 
vanishes. 
 In the on-shell representation, we fix
the position of one scalar impurity and sum over positions of the rest
of the impurities. To explain more explicitly, let us consider the
case of three impurities. In this case, we have
\eqn\on{{\cal O}^J_{\rmon}=\sum_{0\leq l_2,l_3\leq J}
\Tr\left(\phi_{d_1}Z\cdots Z\phi_{d_2}
Z\cdots Z\phi_{d_3}Z\cdots Z\right)
s_2^{l_2}s_3^{l_3},}
where $d_i \in \{1,2,3,4\}$ is the direction of the $i$-th impurity,
$l_i$ is the number\foot{In \ParnachevKK, $l_i$ is argued to include the
number of other impurities in front of it, but the difference is only
subleading in $1/J$ and inconsequential throughout this paper.} of $Z$
in front of $\phi_{d_i}$ and $s_i=e^{2\pi in_i/J}$ is the phase
assigned to $\phi_{d_i}$. This definition gets ambiguous when two
impurities sit next to each other. Therefore, we need a rigorous
definition:
\eqn\onRigor{{\cal O}^J_{\rmon}
={\cal O}^J_{\rmon,\rmc}+{\cal O}^J_{\rmon,\rma},}
with
\eqn\OcOa{\eqalign{
{\cal O}^J_{\rmon,\rmc}
&=\sum_{0\leq a_2\leq a_3\leq J}
\Tr\bigl(\phi_{d_1}Z^{a_2}\phi_{d_2}Z^{a_3-a_2}\phi_{d_3}Z^{J-a_3}\bigr)
s_2^{a_2}s_3^{a_3},\cr
{\cal O}^J_{\rmon,\rma}
&=\sum_{0\leq a_3\leq a_2\leq J}
\Tr\bigl(\phi_{d_1}Z^{a_3}\phi_{d_3}Z^{a_2-a_3}\phi_{d_2}Z^{J-a_2}\bigr)
s_2^{a_2}s_3^{a_3}.}}
To normalize this operator canonically, let us compute its free
two-point function:
\eqn\freeOn{\la\bar{\cal O}^J_{\rmon}{\cal O}^J_{\rmon}\ra
=\la\bar{\cal O}^J_{\rmon,\rmc}{\cal O}^J_{\rmon,\rmc}\ra
+\la\bar{\cal O}^J_{\rmon,\rma}{\cal O}^J_{\rmon,\rma}\ra
=(J+1)(J+2)N^{J+3}.}
Here we have counted the number of pairs $(a_2,a_3)$ such that
$0\leq a_2\leq a_3\leq J$, which is
$\Bigl({\displaystyle J+2\atop\displaystyle 2}\Bigr)$.
Hence, in the BMN limit, the correct normalization is:
\eqn\normOn{{\cal O}^J_{\BMN}={1\over J\sqrt{N^{J+3}}}
{\cal O}^J_{\rmon}.}

Now let us move on to the off-shell representation. In the off-shell
representation, we do not fix the position of any scalar impurity and
treat them on equal footing by summing over all possible positions of
all impurities.
For our present case of 3 impurities, we define
\eqn\off{{\cal O}^J_{\rmoff}=\sum_{0\leq l_1,l_2,l_3\leq J}
\Tr\left(Z\cdots Z\phi_{d_1}Z\cdots Z\phi_{d_2}Z\cdots Z
\phi_{d_3}Z\cdots Z\right)s_1^{l_1}s_2^{l_2}s_3^{l_3},}
where $l_i$ is defined in the same way as above.
Again, a rigorous definition is given by
\eqn\offRigot{{\cal O}^J_{\rmoff}
={\cal O}^J_{\rmoff}(1,2,3)+{\cal O}^J_{\rmoff}(2,3,1)
+{\cal O}^J_{\rmoff}(3,1,2)+{\cal O}^J_{\rmoff}(1,3,2)
+{\cal O}^J_{\rmoff}(3,2,1)+{\cal O}^J_{\rmoff}(2,1,3),}
with
\eqn\Ooff{\eqalign{{\cal O}^J_{\rmoff}(1,2,3)
&=\sum_{0\leq a_1\leq a_2\leq a_3\leq J}
\Tr\left(Z^{a_1}\phi_{d_1}Z^{a_2-a_1}\phi_{d_2}Z^{a_3-a_2}\phi_{d_3}
Z^{J-a_3}\right)s_1^{a_1}s_2^{a_2}s_3^{a_3},\cr
{\cal O}^J_{\rmoff}(2,3,1)
&=\sum_{0\leq a_2\leq a_3\leq a_1\leq J}
\Tr\left(Z^{a_2}\phi_{d_2}Z^{a_3-a_2}\phi_{d_3}Z^{a_1-a_3}\phi_{d_1}
Z^{J-a_1}\right)s_1^{a_1}s_2^{a_2}s_3^{a_3},\cr
&\vdots}}
where the other operators are defined likewise.
Now the claim is that ${\cal O}^J_{\rmoff}$ is non-vanishing if and
only if $n_1+n_2+n_3=0$:
\eqn\claim{{\cal O}^J_{\rmoff}\neq 0\Longleftrightarrow
n_1+n_2+n_3=0.}
Note that this condition is exactly the level-matching condition in
string field theory. Furthermore, if this condition holds, we have
\eqn\OffOn{\eqalign{
{\cal O}^J_{\rmoff}(1,2,3)+{\cal O}^J_{\rmoff}(2,3,1)
+{\cal O}^J_{\rmoff}(3,1,2)&=J{\cal O}^J_{\rmon,\rmc},\cr
{\cal O}^J_{\rmoff}(1,3,2)+{\cal O}^J_{\rmoff}(3,2,1)
+{\cal O}^J_{\rmoff}(2,1,3)&=J{\cal O}^J_{\rmon,\rma},}}
and the off-shell representation \offRigot\ is reduced to the
on-shell one \onRigor:
\eqn\OoffeqJOon{{\cal O}^J_{\rmoff}=J{\cal O}^J_{\rmon}.}
This explains the terminology of ``on-shell/off-shell'' representation.
Consequently, the correct normalization of the off-shell operator is
\eqn\normOff{{\cal O}^J_{\BMN}
={1\over\sqrt{J}\sqrt{J^3}\sqrt{N^{J+3}}}{\cal O}^J_{\rmoff}.}

This argument can be immediately generalized to $n$ impurities
assuming all of them are different. The on-shell operator is
\eqn\nOn{{\cal O}^{J}_{\rmon}=\sum_{0\leq l_2,\cdots,l_n\leq J}
\Tr\left(\phi_{d_1}Z\cdots Z\phi_{d_2} Z\cdots\cdots Z\phi_{d_n}
Z\cdots Z\right)\prod_{i=2}^ns_i^{l_i},}
with $l_i$ being the number of $Z$'s in front of $\phi_i$ as before.
Or more rigorously the definition of it is given as the sum of
$(n-1)!$ operators corresponding to permutations after fixing the
position of one impurity:
\eqn\nOnRigor{{\cal O}^{J}_{\rmon}
=\sum_{\sigma\in{\rm Perm}\{2,\cdots,n\}}{\cal O}^{J}_{\rmon,\sigma},}
with
\eqn\nOnSigma{{\cal O}^{J}_{\rmon,\sigma}
=\sum_{0\leq a_{\sigma(2)}\leq\cdots\leq a_{\sigma(n)}\leq J}
\Tr\left(\phi_{d_1}Z^{a_{\sigma(2)}}\phi_{d_{\sigma(2)}}
Z^{a_{\sigma(3)}-a_{\sigma(2)}}\phi_{d_{\sigma(3)}}
\cdots\phi_{d_{\sigma(n)}}Z^{J-a_{\sigma(n)}}\right)
\prod_{i=2}^ns_i^{a_i}.}
Each ${\cal O}^{J,n}_{on,\sigma}$ is composed of
$\Bigl({\displaystyle J+n-1\atop\displaystyle n-1}\Bigr)
\simeq {\displaystyle J^{n-1}\over\displaystyle(n-1)!}$ terms,
where the combinatoric number comes from the number of $(n-1)$-tuple
$(a_2,a_3,\cdots,a_{n})$ satisfying
$0\leq a_{\sigma(2)}\leq\cdots\leq a_{\sigma(n)}\leq J$.
Hence, the normalization is
\eqn\normnOn{{\cal O}^{J}_{\BMN}={1\over\sqrt{J^{n-1}}\sqrt{N^{J+n}}}
{\cal O}^{J}_{\rmon}.}
Similarly, the off-shell operator with $n$ impurities is
\eqn\nOff{{\cal O}^{J}_{\rmoff}=\sum_{0\leq l_1,\cdots,l_n\leq J}
\Tr\left(Z\cdots Z\phi_{d_1} Z\cdots Z\phi_{d_2} Z\cdots\cdots
Z\phi_{d_n}Z\cdots Z\right)\prod_{i=1}^ns_i^{l_i},}
with a rigorous definition given by a sum over $n!$ terms.
As in 3-impurity case, we have
\eqn\nOffnOn{{\cal O}^{J}_{\rmoff}=J{\cal O}^{J}_{\rmon},}
if and only if the level-matching (on-shell) condition holds for the
off-shell operator.
Hence the normalization for the off-shell operator is
\eqn\normnOn{{\cal O}^{J}_{\BMN}
={1\over\sqrt{J}\sqrt{J^n}\sqrt{N^{J+n}}}{\cal O}^{J}_{\rmoff}.}
Here we can think of each impurity as carrying a normalization factor
$1/\sqrt{J}$, since we sum over $J$ possible positions for each
impurity.
The leftover factor $1/\sqrt{J}$ is the original normalization of the
vacuum operator and it originates in the cyclic property of $\Tr$.

So far, we have defined on-shell and off-shell operators assuming that
all impurities are distinct. However, we have to deal with same
impurities eventually since there are only $4$ directions. When two
impurities, say $\phi_{d_1}$ and $\phi_{d_2}$, are the same, we have
to insert $-\bar{Z}$ when they collide as discussed in Appendix G.
Then the correct definition is in the off-shell representation,
\eqn\nOffSame{\eqalign{
&{\cal O}^{J}_{\rmoff}=\sum_{0\leq l_1,l_2,l_3,\cdots,l_n\leq J}
\Tr\left(Z\cdots Z\phi_{d_1}Z\cdots Z\phi_{d_2}Z\cdots Z\phi_{d_3}
Z\cdots\cdots Z\phi_{d_n}Z\cdots Z\right)
\prod_{i=1}^ns_i^{l_i}\cr
&\quad-\sum_{0\leq l_{(1,2)},l_3,\cdots,l_n\leq J+1}
\Tr\left(Z\cdots Z\bar{Z}Z\cdots Z\phi_{d_3}
Z\cdots\cdots Z\phi_{d_n}Z\cdots Z\right)
(s_1s_2)^{l_{(1,2)}}\prod_{i=3}^ns_i^{l_i},}}
where $l_{(1,2)}$ is the number of $Z$'s in front of $\bar{Z}$ arising
when $\phi_{d_1}$ and $\phi_{d_2}$ collide. Now we have to do this
modification whenever we have a pair $(i,j)$ such that $d_i=d_j$.
However, as argued in Appendix G, we do not have to worry about
collision of more than two impurities.
The normalization is not changed since the number of $\bar{Z}$ terms
is subleading in $1/J$ compared with the original terms because
$\bar{Z}$ terms arise only when two impurities collide.

As an example, let us consider a BMN operator with 4 same impurities,
i.e. $d_1=d_2=d_3=d_4$. In this case the BMN operator should be
modified by $-{\bar Z}$ as
\eqn\fourOffSame{\eqalign{
&{\cal O}^{J}_{\rmoff}=\sum_{0\leq l_1,l_2,l_3,l_4\leq J}
\Tr\left(Z\cdots Z\phi_{d_1}Z\cdots Z\phi_{d_2}Z\cdots Z\phi_{d_3}
Z\cdots Z\phi_{d_4}Z\cdots Z\right)
\prod_{i=1}^4s_i^{l_i}\cr
&\quad-\sum_{0\leq l_{(1,2)},l_3,l_4\leq J+1}
\Tr\left(Z\cdots Z\bar{Z}Z\cdots Z\phi_{d_3}Z\cdots Z\phi_{d_4}
Z\cdots Z\right)(s_1s_2)^{l_{(1,2)}}s_3^{l_3}s_4^{l_4}\cr
&\quad-\sum_{0\leq l_{(1,3)},l_2,l_4\leq J+1}
\Tr\left(Z\cdots Z\bar{Z}Z\cdots Z\phi_{d_2}Z\cdots Z\phi_{d_4}
Z\cdots Z\right)(s_1s_3)^{l_{(1,3)}}s_2^{l_2} s_4^{l_4} \cr
&\qquad\qquad\qquad\qquad\vdots\cr
&\quad+\sum_{0\leq l_{(1,2)},l_{(3,4)}\leq J+2}
\Tr\left(Z\cdots Z\bar{Z}Z\cdots Z\bar{Z}Z\cdots Z\right)
(s_1s_2)^{l_{(1,2)}}(s_3s_4)^{l_{(3,4)}}\cr
&\qquad\qquad\qquad\qquad\vdots}}

\appendix{I}{The ${\cal O}(g_2)$ coupling an $p$-th string state to an
$p+1$-th string state}

In a recent paper Gursoy \GursoyFJ\ has analyzed the two-point
function of multi-trace BMN operators with two different
impurities. The class of operators he considers are:
\eqn\multitracemany{
{\cal T}^{J,y_1,y_2,\ldots,y_{p}}_{ij,n}
=:{\cal O}_n^{y_1\cdot J}\cdot{\cal O}^{y_2\cdot J}\cdot\cdot\cdot
{\cal O}^{y_{p}\cdot J}:\delta_{y_1+\ldots +y_p,1}.}
The ${\cal O}(g_2)$ result for the mixing of the $p$-th trace with the
$p+1$ trace BMN operator is given by \GursoyFJ\
\eqn\mixingtrace{\eqalign{
G^{{p,p+1}(1)}_{ny_1\ldots y_p;mz_1\ldots z_p}
&=y_1^{3/2}C_{n,mz_1/y_1}\sum_P \delta_{y_2,z_{P(2)}}\ldots
\delta_{y_p,z_{P(p)}}
\delta_{y_1,z_{P(p)}}\delta_{y_1,z_1+z_{P(i+1)}}+\cr
&{1\over
J}\delta_{n,m}\delta_{y_1,z_1}
\sum_{P,P^\prime}\delta_{y_{P(2)},z_{P^\prime(2)}}\ldots
\delta_{y_{P(p-1)},z_{P^\prime(p-1)}}\delta_{y_{P(p)},z_{P^\prime(p)}+
z_{P^\prime(p+1)}}.}}
The contribution in the first line is due to contractions in which the
two impurities in the $p$-trace operator contract with the two
impurities and any vacuum operator in the $p+1$-trace operator.
Therefore, the quantity $C_{n,mz_1/y_1}$ is the mixing between a
single trace and double trace two-impurity BMN operator. The
contribution in the second line comes from Wick contractions where the
operators with the impurities just connect among themselves and the
vacuum operators in the $p$-th trace BMN operator contract with the
vacuum operator in the $p+1$-th trace BMN operator.

The contribution to the the matrix of anomalous dimensions is given by
\GursoyFJ
\eqn\anomtrace{
\Gamma^{{p,p+1}(1)}_{ny_1\ldots y_p;mz_1\ldots
z_p}=\left( {n^2\over y_1^2}+{m^2\over z_1^2}-{nm\over
y_1z_1}\right)G^{{p,p+1}(1)}_{ny_1\ldots y_p;mz_1\ldots
z_p}.}
Using the holographic proposal we can calculate these matrix elements
in the orthonormal basis:
\eqn\orthonor{\eqalign{
{\tilde \Gamma}^{{p,p+1}(1)}_{ny_1\ldots y_p;mz_1\ldots
z_p}&=\Gamma^{{p,p+1}(1)}_{ny_1\ldots y_p;mz_1\ldots
z_p}-{1\over 2}\left\{G^{{p,p+1}(1)}_{ny_1\ldots y_p;mz_1\ldots
z_p},\Gamma^{{p,p+1}(0)}_{ny_1\ldots y_p;mz_1\ldots
z_p}\right\}\cr
&={1\over 2}\left({n\over y_1}-{m\over
z_1}\right)^2G^{{p,p+1}(1)}_{ny_1\ldots y_p;mz_1\ldots
z_p}.}}
We note that in the orthonormal basis that the second term in
\mixingtrace\ does not contribute due to the $\delta$ function
constraints and the prefactor in \orthonor . Therefore,
the final answer can be
written as:
\eqn\finall{
{\tilde \Gamma}^{{p,p+1}(1)}_{ny_1\ldots y_p;mz_1\ldots
z_p}={1\over \sqrt{y_1}}{\tilde \Gamma}_{n,m
z_1/y_1}\times\sum_{P}\delta_{y_2,z_{P(2)}}\ldots
\delta_{y_p,z_{P(p)}}
\delta_{y_1,z_{P(p)}}\delta_{y_1,z_1+z_{P(i+1)}}.}

We now perform the relevant string field theory calculation. The
string states dual to the BMN operators \multitracemany\ are given by
\eqn\stringstatesdual{
|n,y_1,y_2,\ldots ,y_p\ra=\alpha^{i}_n\alpha^{j}_
{-n}|\hbox{vac},y_1\ra\otimes
       |\hbox{vac},y_2\ra\otimes \ldots\otimes  |\hbox{vac},y_p\ra\
\delta_{y_1+\ldots +y_p,1}.}
It follows from the expression for the cubic Hamiltonian vertex
\Hone\Prefa\ that any contraction coupling only vacua is zero. The
only possible non-zero contributions are those in which the string
carrying the two impurities contracts with the string carrying
two-impurities and a vacuum string state. Therefore, the matrix
elements of the unitly normalized states are:
\eqn\matrixHamil{\eqalign{
&{1\over \mu}\la n, y_1\ldots y_p|H_3|m, z_1\ldots z_p \ra\cr
&={1\over \mu}\la n, y_1|H_3|m, z_1\ra\times
\sum_{P}\delta_{y_2,z_{P(2)}}\ldots
\delta_{y_p,z_{P(p)}}
\delta_{y_1,z_{P(p)}}\delta_{y_1,z_1+z_{P(i+1)}}\cr
&={1\over \sqrt{y_1}}{\tilde
\Gamma}^{(1)}_{n,mz_1/y_1}\times\sum_{P}\delta_{y_2,z_{P(2)}}\ldots
\delta_{y_p,z_{P(p)}}
\delta_{y_1,z_{P(p)}}\delta_{y_1,z_1+z_{P(i+1)}},}}
which match the gauge theory computation.

\listrefs
\bye